\documentclass[preprint,12pt]{elsarticle}

\usepackage{amssymb}
\usepackage{amsmath}
\usepackage{float}
\usepackage{subcaption}
\usepackage{graphicx}
\usepackage{booktabs}
\usepackage{array}
\usepackage{siunitx}
\usepackage{booktabs}
\usepackage{array}
\usepackage{makecell}
\usepackage{caption}

\journal{Journal of Computational Physics}

\begin{document}

\begin{frontmatter}



\title{A volume penalization method for solving conjugate scalar transport with interfacial jump conditions}


\author{Ming Liu} 
\author{Yosuke Hasegawa\corref{cor1}} 
\cortext[cor1]{Corresponding author.}

\affiliation{organization={Center for Research on Innovative Simulation Software, Institute of Industrial Science, The University of Tokyo},
            addressline={4-6-1 Komaba, Meguro-ku}, 
            city={Tokyo},
            postcode={153-8505}, 
            state={},
            country={Japan}}

\begin{abstract}
Conjugate scalar transport with interfacial jump conditions on complex interfacial geometries is common in thermal and chemical processes, while its accurate and efficient simulations are still quite challenging. In the present study, a novel treatment of a two-phase interface in the volume penalization method, a kind of immersed boundary method, for solving conjugate scalar transport with general interfacial boundary conditions is developed. We first propose an interfacial treatment for solving an advection-diffusion equation with a Neumann boundary condition, and then extend it to general conjugate scalar transport with both interfacial flux and scalar jumps. A one-dimensional diffusion problem is solved to verify the present scheme and demonstrate the advantage of the present scheme in improving accuracy and unifying the governing equations in the two phases with an additional source term representing the local jump condition of the interfacial scalar flux. Then, the present scheme is further applied to fluid-solid coupled scalar diffusion and advection-diffusion problems with the scalar and its flux jumps across the interface. The simulation results of the present scheme generally show good agreement with reference results obtained by body-fitted mesh simulations with average relative deviations less than 3.0\%.
\end{abstract}


\end{frontmatter}



\section{Introduction}
\label{sec1}

Conjugate scalar transport in different phases is often encountered in various industrial processes. At the interface of different phases, a jump of scalar or its flux may exist, and it cannot be neglected in some applications. Typical examples include a heat flux jump at a gas-solid interface in microchannels \cite{Chibouti21}, a scalar flux jump at an interface due to chemical reactions \cite{ZouL18}, a temperature jump at a fluid-solid interface in microscale heat transfer systems due to the non-equilibrium effect \cite{GuoY16}, and an interfacial concentration jump in gas-liquid systems \cite{ZhangZ18}. Although conjugate scalar transport phenomena are widespread in thermal and chemical engineering practice, their effective and accurate simulations are still challenging due to the difficulty in treating interfacial jump conditions, especially for complex geometries.

For numerical simulations using a body-fitted mesh, different phases are usually simulated separately and then coupled by explicitly imposing interfacial boundary conditions. Zhang et al. \cite{ZhangL10a}\cite{ZhangL10b} performed simulations of conjugate heat and mass transfer in a membrane-based heat exchanger. An interfacial jump condition is then further considered in their air-membrane-solution system \cite{ZhangL12}, where a heat flux jump at the gas-liquid interface due to the latent heat flux exists, and an alternating direction implicit technique is employed to solve the coupling problem. Of particular interest for the past few years is the application of the Immersed Boundary Method (IBM) to complex geometries \cite{Verzicco23}. Since IBM allows to embed an arbitrary interfacial geometry in a Cartesian grid system, it does not require cumbersome grid generation for each geometry. Meanwhile, the grid points are not generally located on the interface, and therefore, there exists an inherent difficulty in imposing interfacial boundary conditions accurately. Leveraging the advantage of the lattice Boltzmann method (LBM) in dealing with complex boundaries, the immersed-boundary LBM has been developed for conjugate scalar transport with general interfacial conditions by several scholars. Hu et al. \cite{HuZ16}\cite{HuZ17} proposed a scheme to treat general interfacial boundary conditions with and without jumps \cite{HuangJ15}. Mu et al. \cite{MuY18} developed an interfacial treatment for conjugate scalar transport problems with Dirichlet and Neumann boundary conditions \cite{LiL14}. While the second-order accuracy for flat interfaces and the first-order accuracy for curved ones can be achieved by the immersed-boundary LBM \cite{Korba20}, the computational costs would rise considerably due to complicated interpolation-based treatments around the interface. In contrast, immersed boundary schemes for conjugate scalar transport under a finite volume framework are relatively limited. Kumar and Natarajan developed a diffuse-interface immersed-boundary scheme based on the volume of body function to solve conjugate heat transfer problems \cite{Kumar19}, where unified governing equations in fluid and solid regions are solved simultaneously \cite{PanD06}. However, the above schemes are applicable only to conjugate heat transfer problems without interfacial jumps. Besides, for conjugate heat transfer problems involving phase change materials, the enthalpy-based method \cite{Blais18}\cite{Ahn23} that incorporates the latent heat into the total enthalpy is commonly adopted to consider the jump in the enthalpy during the phase change. Although the enthalpy-based method can effectively incorporate a prescribed heat flux jump due to the latent heat associated with phase change, it is difficult to handle arbitrary jumps in the local scalar flux Hence, the immersed boundary scheme to treat general interfacial jump conditions under the framework of the finite volume method remains to be developed.

To this end, the Volume Penalization Method (VPM), which is a typical diffuse-interface IBM, is introduced in the present study. Inspired by numerical models of porous media, the VPM was originally proposed by Schneider et al. \cite{Schneider05}\cite{Kadoch12} to model solid regions as porous media with a sufficiently small permeability, where an artificial source term is added to force the local fluid velocity approaches to zero inside the solid region. Initially, the VPM is employed to impose a Dirichlet boundary condition, especially a no-slip velocity boundary condition \cite{Kadoch16}\cite{Kametani20}. Then, it has been further extended to deal with the Neumann \cite{Sakurai19}\cite{Kolomenskiy15} and Robin conditions \cite{Thirumalaisamy22}. Generally, the Neumann boundary condition in VPM is imposed by adding a source term, so that the derivative of a variable of interest approaches to a target value \cite{Brown14}. However, since the penalization acts inside the entire volume of the solid, the existing scheme also generates a non-physical solution in the solid region. Therefore, it is not suitable for solving scalar transfer problems with coupled fluid and solid phases. In addition, a numerical method for dealing with interfacial jump conditions in VPM has not been established yet.

The objective of the present work is to develop a novel interfacial treatment in a VPM for the Neumann boundary condition and also general interfacial conditions with jumps in scalar and its flux. This paper is organized as follows: The VPM for conjugate scalar transport with an interfacial jump condition is proposed in Section \ref{sec2}. In Section \ref{sec3}, a one-dimensional diffusion problem is employed to verify the treatment for the Neumann boundary condition, and the present VPM is compared with the existing scheme. In Section \ref{sec4} and Section \ref{sec5}, more general coupled fluid-solid scalar diffusion and advection-diffusion problems are simulated using the present VPM, respectively, and the results are compared with those of simulations with the body-fitted method (BFM). Finally, the conclusions of the present study are summarized in Section \ref{sec6}.

\section{Volume penalization method}
\label{sec2}
\subsection{Problem description}
\label{subsec2_1}
\begin{figure}[H]
\centering
\includegraphics[width=0.8\textwidth]{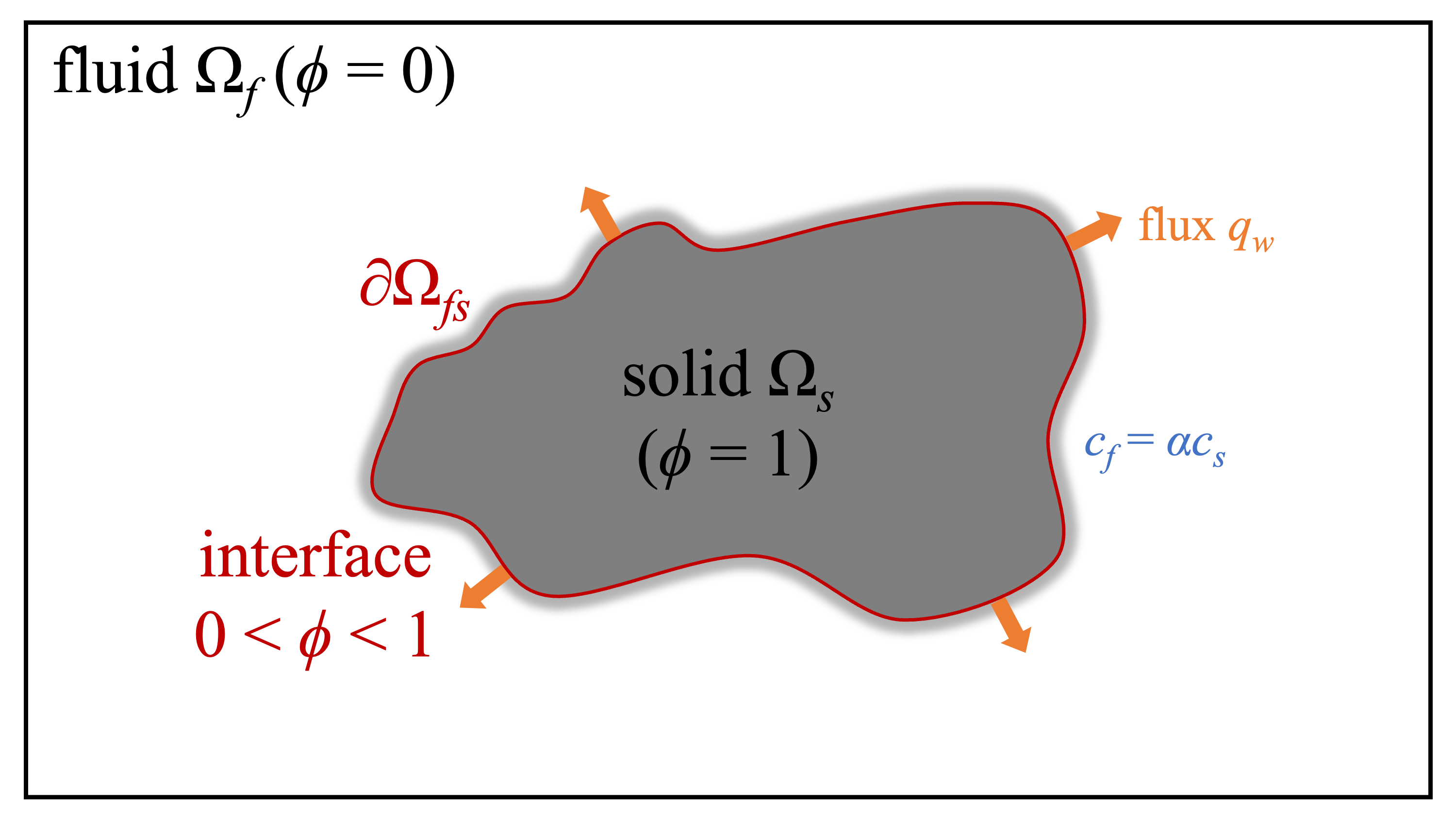}
\caption{Schematic of conjugate scalar transport with interfacial jump conditions.}\label{fig1}
\end{figure}

A conjugate transport process of a scalar $\mathit{c}$, e.g., concentration or temperature, coupled in fluid-solid regions $\Omega$, is considered in the present study, as illustrated in Fig. \ref{fig1}. Throughout the manuscript, the subscripts $\mathit{f}$ and $\mathit{s}$ denote quantities in the fluid and solid regions, respectively. We assume a local jump in scalar flux $\mathit{q_w}$ at the fluid-solid interface $\partial \Omega_{\mathit{fs}}$, which may be caused by chemical reactions or latent heat arising from evaporation or condensation, and this additional scalar flux leads to a scalar flux jump at the fluid-solid interface. The discontinuity of the scalar itself at the fluid-solid interface is also considered here. Specifically, we assume a scalar jump at the interface in the form of a constant multiplier $\alpha$, namely, $\mathit{c}_{f} = \alpha\,\mathit{c}_{s}$. As an example, $\alpha$ corresponds to the ratio of heat capacity between fluid and solid, where $\mathit{c}$ is the enthalpy in a conjugate heat transfer system. Also, in the case of mass transfer across a gas-liquid interface, $\alpha$ corresponds to the Henry constant for a solute concentration $\mathit{c}$.

The governing equations and the corresponding interfacial boundary conditions to describe the above conjugate scalar transport problem are the advection-diffusion equation in the fluid region and the diffusion equation in the solid region, as shown below:

fluid region:
\begin{equation}
\frac{\partial c_f}{\partial t}
+ \boldsymbol{u} \cdot \nabla c_f
= \nabla \cdot \left( D_f \nabla c_f \right),
\end{equation}

solid region:
\begin{equation}
\frac{\partial c_s}{\partial t}
= \nabla \cdot \left( D_s \nabla c_s \right).
\end{equation}

The boundary condition at the interface is expressed as
\begin{equation}
\left\{
\begin{aligned}
c_f &= \alpha_s c_s, \\
- D_s \frac{\partial c_s}{\partial n} + q_w
&= - D_f \frac{\partial c_f}{\partial n},
\end{aligned}
\right.
\qquad \text{on } \partial \Omega_{fs} ,
\end{equation}
where $\mathit{c}$, $\boldsymbol{u}$, $\mathit{D}$, $\boldsymbol{n}$ are a scalar, a fluid velocity vector, a molecular diffusivity of the scalar, and the wall-normal vector pointing outward from the solid to the surrounding fluid, respectively.

\subsection{Level-set function}
\label{subsec2_2}
\begin{figure}[H]
\centering
\includegraphics[width=0.8\textwidth]{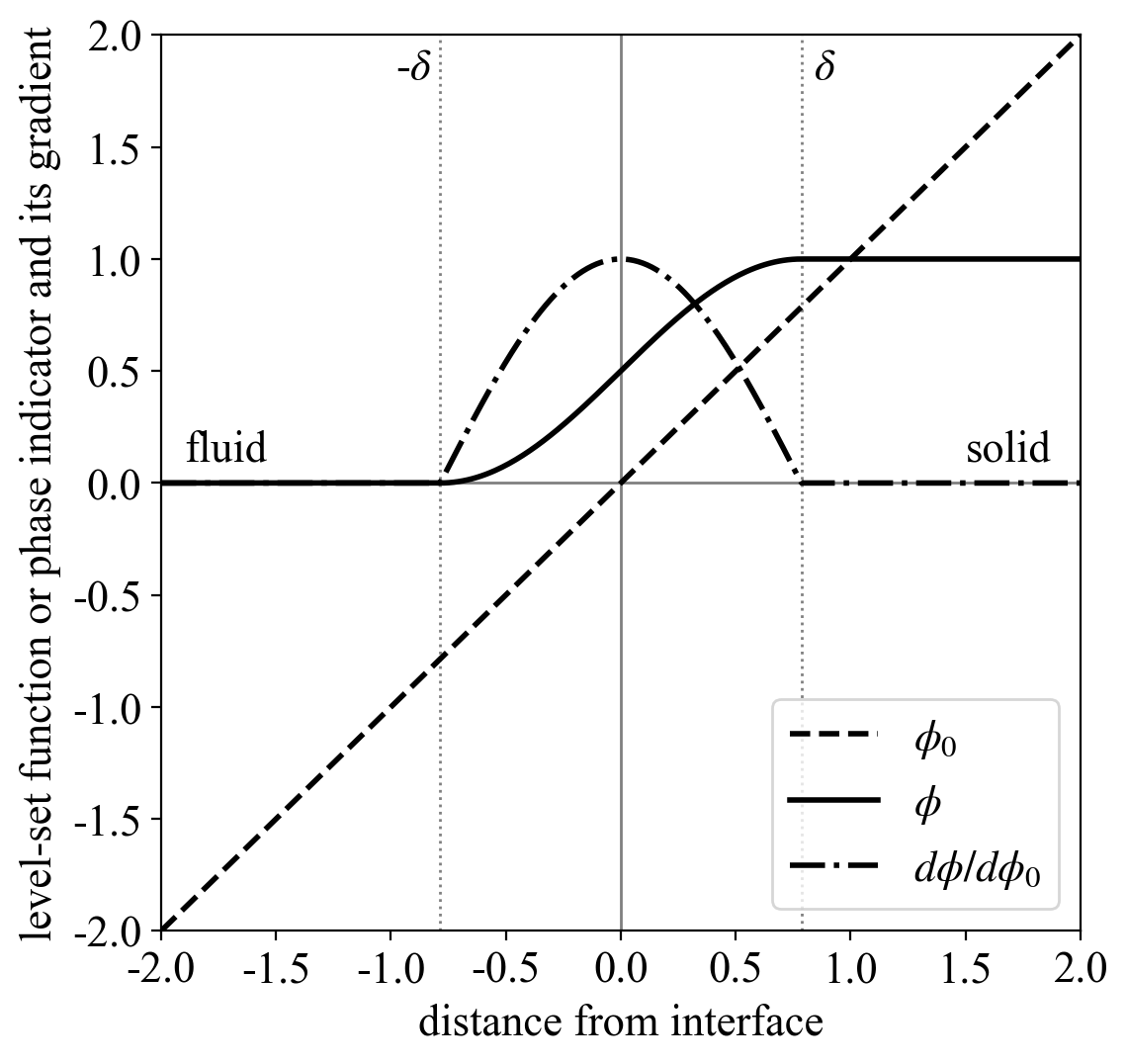}
\caption{Profiles of the level-set function $\phi_{0}$, the phase indicator $\phi$, and its gradient $\frac{d\phi}{d\phi_0}$ in the vicinity of the fluid-solid interface.}\label{fig2}
\end{figure}

In the present study, the level-set function $\phi_{0}$ is employed to represent an arbitrary three-dimensional fluid-solid interface embedded in a Cartesian grid system. The level-set function is defined as a signed distance from the fluid-solid interface. The value of the level-set function is set to be zero on the interface, positive in the solid region $\Omega_{s}$, and negative in the fluid region $\Omega_{f}$. The distribution of the local level-set function $\phi_{0}$ can be converted into the phase indicator $\phi$ by the following formula, so that $\phi$ = 0 in the fluid region, $\phi$ = 1 in the solid region, and 0 < $\phi$ < 1 in an intermediate interfacial region.
\begin{equation}
\phi =
\begin{cases}
0, & \phi_0 < -\delta, \\[6pt]
\dfrac{1}{2}\sin\!\left( \dfrac{\phi_0}{\delta}\,\dfrac{\pi}{2} \right) + \dfrac{1}{2},
& -\delta \le \phi_0 \le \delta, \\[10pt]
1, & \phi_0 > \delta ,
\end{cases}
\end{equation}
where $\delta$ is the half width of an interfacial region (0 < $\phi$ < 1) between the fluid and solid regions. By means of the level-set function, the fluid-solid interface can be represented by a diffused interfacial region with a thickness of 2$\delta$. In general, the interfacial half-width $\delta$ is set to be the width of a few local grid points. Therefore, the interfacial region becomes narrower by reducing the local grid size, and resultant numerical results are expected to converge to the solutions for an infinitesimal interfacial thickness. In the present simulations, the interfacial thickness is determined by the local diagonal grid spacing multiplied by a constant coefficient $\mathit{K}_{\delta}$, which is set as 1.5 in the present study. Figure \ref{fig2} schematically shows the profiles of the level-set function $\phi_{0}$ and the corresponding phase indicator $\phi$ near a fluid-solid surface. The gradient of the phase indicator, $\frac{d\phi}{d\phi_0}$, is also plotted, whose integration across the interfacial region is unity by its definition. This property will be used later to achieve a given jump condition in the local scalar flux across the interface.

\subsection{VPM for Neumann condition}
\label{subsec2_3}
In order to develop a VPM for a conjugate scalar transport problem with interfacial jump conditions, one of the most basic conditions, i.e., a Neumann boundary condition in a single fluid phase, is considered as a first step. The advection-diffusion equation for scalar transport with a Neumann boundary condition can be expressed as follows:
\begin{equation}
\frac{\partial c_f}{\partial t}
+ \boldsymbol{u} \cdot \nabla c_f
= \nabla \cdot \left( D_f \nabla c_f \right),
\end{equation}
\begin{equation}
\boldsymbol{q}_w
= - D_f \frac{\partial c_f}{\partial n}
= - D_f \left( \nabla c_f \cdot \boldsymbol{n} \right) \boldsymbol{n},
\qquad \text{on } \partial \Omega_f .
\end{equation}
The wall scalar flux $\boldsymbol{q}_{w}$ from the wall to the fluid can be expressed by the product of the wall normal vector $\boldsymbol{n} = -\,\frac{\nabla \phi_0}{\lvert \nabla \phi_0 \rvert}$ pointing outward from the solid to the surrounding fluid and its intensity $\mathit{q_w}$ as
\begin{equation}
\boldsymbol{q}_w = q_w\,\boldsymbol{n}.
\end{equation}
Note that it is equivalent to the scalar flux jump as defined in Sec. \ref{subsec2_1}, assuming that $\mathit{c}_{s}$ = 0 throughout the solid region.

In the existing VPM \cite{Brown14} for a Neumann boundary condition, its governing equation can be written as
\begin{equation}
\frac{\partial c}{\partial t}
+ \boldsymbol{u} \cdot \nabla c
=
\nabla \cdot \left( D_f \nabla c \right)
- \eta\, \phi
\left( D_f \nabla c \cdot \boldsymbol{n} + q_w \right),
\label{eq8}
\end{equation}
where the scalar flux $-D_f \nabla c \cdot \boldsymbol{n}$ is forced to approach a target magnitude $\mathit{q_w}$ through the VPM term, appearing as the last term on the right-hand side of Eq. (\ref{eq8}). By means of VPM, both fluid and solid regions are solved with the unified governing equation (\ref{eq8}). Due to the VPM term, a Neumann boundary condition is achieved in the fluid region, while a non-physical scalar field is also generated and diffused in the solid region.

\begin{figure}[H]
\centering
\begin{subfigure}{0.48\linewidth}
    \caption{}
    \includegraphics[width=\linewidth]{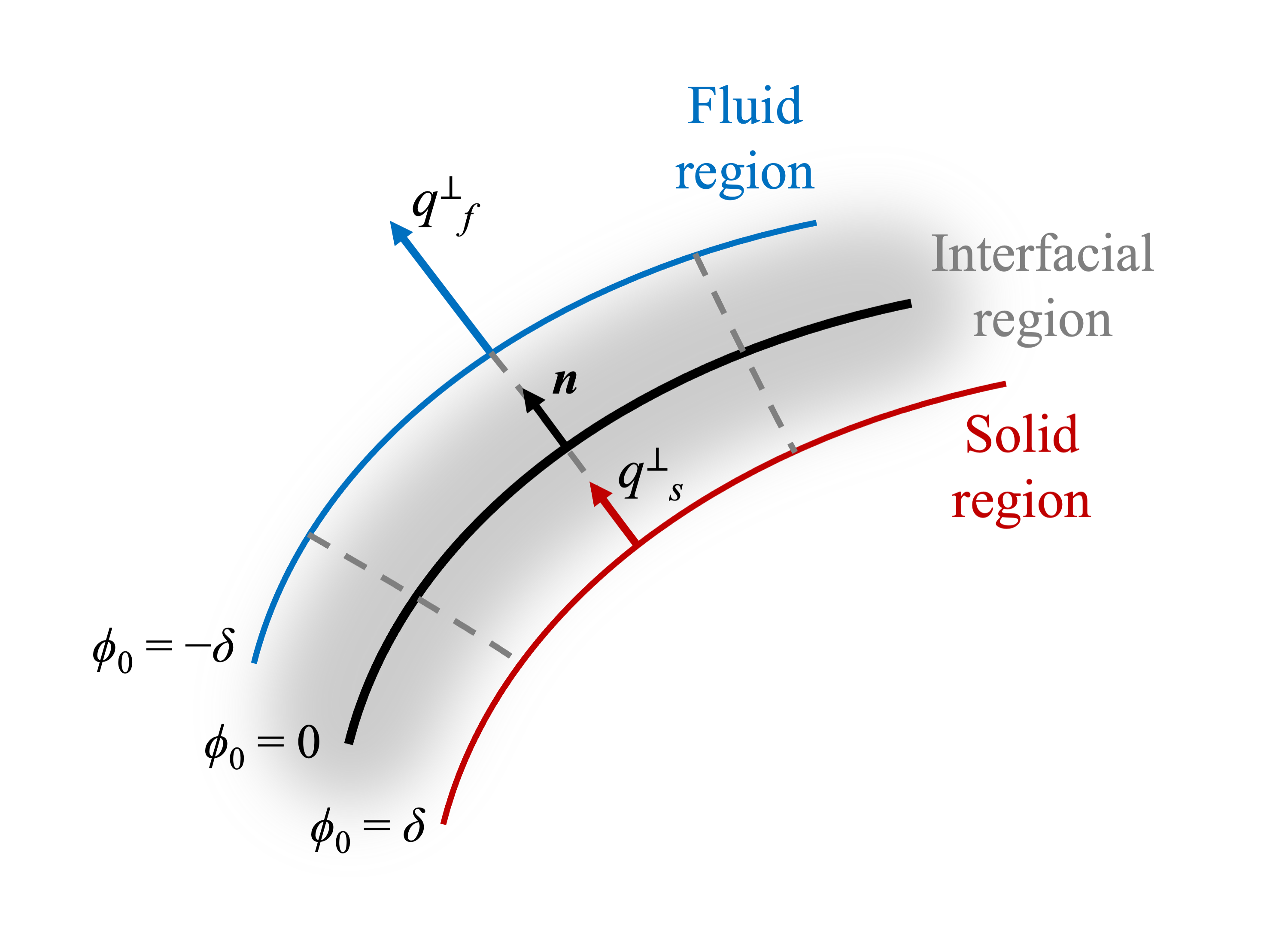}
    \centering
\end{subfigure}
\hfill
\begin{subfigure}{0.48\linewidth}
    \caption{}
    \includegraphics[width=\linewidth]{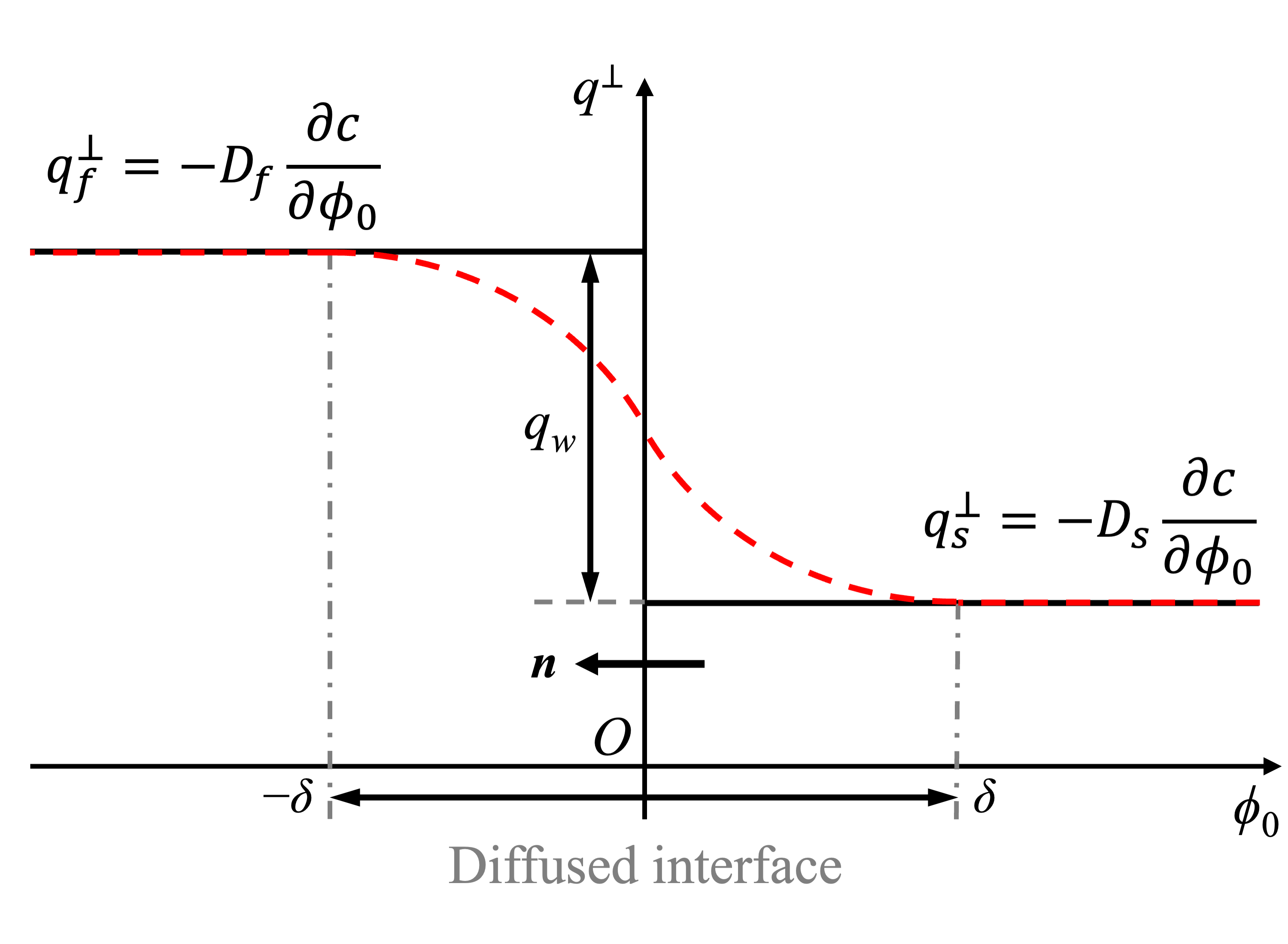}
    \centering
\end{subfigure}

\caption{(a) Schematic and (b) profile of the distribution of the scalar flux around the interface.}
\label{fig3}
\end{figure}

Figures \ref{fig3} (a) and (b) illustrate the schematic of the distribution of the scalar flux around the interface and its profile in the wall-normal direction pointing outward from the solid to the fluid, respectively. The wall-normal scalar flux $q^{\perp}$ with a jump condition can be written as
\begin{equation}
q^{\perp}(\phi_0)
=
q_f^{\perp}
- \bigl( q_f^{\perp} - q_s^{\perp} \bigr) H(\phi_0)
=
q_f^{\perp}
- q_w \, H(\phi_0),
\label{eq9}
\end{equation}
where the subscripts $\mathit{f}$ and $\mathit{s}$ represent quantities on the fluid and solid regions, respectively. $\mathit{H(\phi_0)}$ is the Heaviside step function. Generally, a scalar jump in the wall-normal direction across the interface results in the presence of a source term in the governing equation as

\begin{equation}
q_f^{\perp} - q_s^{\perp}
= q^{\perp}(\phi_0 = 0^{-}) - q^{\perp}(\phi_0 = 0^{+})
= \int_{0^{+}}^{0^{-}} \frac{\partial q^{\perp}}{\partial \phi_0}\, d\phi_0.
\label{eq10}
\end{equation}

In the present diffused-interface VPM, the step change of $q^{\perp}$ across the interface $\phi_{0} = 0$ (see black solid line in Fig. \ref{fig3}(b)) is approximated by the smooth transition from $\phi_{0} = -\delta$ to $\delta$ (see red dashed line in Fig. \ref{fig3}(b)). Therefore, the resultant source term $\frac{\partial q^{\perp}}{\partial \phi_0}$ in the scalar transport equation due to the scalar flux jump can be further derived as

\begin{equation}
\frac{\partial q^{\perp}}{\partial \phi_0}
= \frac{\partial \bigl(- q_w H \bigr)}{\partial \phi_0}
= - q_w \, \delta(\phi_0)
\approx \nabla \cdot \bigl( q_w \, \phi \, \boldsymbol{n} \bigr).
\label{eq11}
\end{equation}
Here, $\delta(x)$ is the one-dimensional delta function that changes across the fluid-solid interface. This delta function is then approximated by $-\nabla \cdot \bigl(\phi \, \boldsymbol{n} \bigr)$ under the assumption that the interfacial thickness is sufficiently small. The negative sign arises because we define the wall normal vector as the opposite of the gradient of the level-set function. Note that $-\nabla \cdot \bigl(\phi \, \boldsymbol{n} \bigr)$ converges to the delta function $\delta(\phi_0)$  as the grid spacing reduces near the interface, and thus the interfacial region becomes narrower. In other words, the present scheme introduces a source term within the thin interfacial region, so that its integral across the interface corresponds to the local scalar flux jump in the wall-normal direction.

Accordingly, the newly proposed VPM equation for an interfacial scalar flux jump $\mathit{q_w}$ can be written as follows:
\begin{equation}
\frac{\partial c}{\partial t}
+ \boldsymbol{u} \cdot \nabla c
=
\nabla \cdot ( D \nabla c )
+ \nabla \cdot \bigl( q_w \, \phi \, \boldsymbol{n} \bigr).
\label{eq12}
\end{equation}
Here, the molecular diffusivity for a scalar is defined in each phase as
\begin{equation}
D(\phi_0)
=
\begin{cases}
D_f, & \phi_0 < 0, \\
D_s = 0, & \phi_0 \ge 0 .
\end{cases}
\label{eq13}
\end{equation}
Note that the last VPM term newly appears on the RHS of Eq. (\ref{eq12}) for an interfacial jump of scalar flux has a non-zero value only in the interfacial region, so that the original governing equation can be reproduced in both fluid and solid regions.

One unique feature of the present VPM term Eq. (\ref{eq11}) is that it is in a divergence form, so that its integral across the interface can be precisely controlled. Specifically, when it is integrated from $\phi_{0} = -\delta$ to $\delta$, we get
\begin{align}
0
&=
\int_{-\delta}^{\delta}
\bigl[
\nabla \cdot ( D \nabla c )
+ \nabla \cdot ( q_w \, \phi \, \boldsymbol{n} )
\bigr]
\, d\phi_0
\notag
\\[6pt]
&=
\bigl( - D_f \nabla c \cdot \boldsymbol{n} \bigr)_{\phi_0 = -\delta}
-
\bigl( - D_s \nabla c \cdot \boldsymbol{n} \bigr)_{\phi_0 = \delta}
+
\bigl[ q_w \phi(-\delta) - q_w \phi(\delta) \bigr]
\notag
\\[6pt]
&=
q_f^{\perp} - q_s^{\perp} - q_w .
\label{eq14}
\end{align}
This shows that the scalar flux jump of $\mathit{q_w}$ can always be realized across the interfacial region.

The advantages of the present VPM in Eqs. (\ref{eq12})-(\ref{eq13}) over the conventional scheme in Eq. (\ref{eq8}) are two folds. First, the prescribed scalar flux $\mathit{q_w}$ can be precisely applied, since the volume integral of $-\nabla \cdot \bigl(\phi \, \boldsymbol{n} \bigr)$ is unity by definition. Second, the molecular diffusivity is set to be zero inside the solid, so that the applied scalar flux is only diffused into the fluid.

\subsection{VPM for general conjugate scalar transport}
\label{subsec2_4}
Next, we extend the present VPM to the conjugate scalar transport problems with an interfacial jump condition. In order to establish a unified governing equation for both fluid and solid regions, an equivalent scalar $\mathit{h}$ from the original scalar $\mathit{c}$ is defined as follows,
\begin{equation}
h = \alpha c,
\end{equation}
where the equivalent coefficient $\alpha$ is equal to 1 in the fluid region and $\alpha_s$ in the solid region. By introducing the equivalent scalar, the original scalar jump at the fluid-solid interface is removed, making it feasible to solve the coupled fluid-solid regions by a unified governing equation. Accordingly, the boundary condition at the fluid-solid interface can be rewritten into the following form:
\begin{equation}
\left\{
\begin{aligned}
h_f &= h_s, \\
- D_s \frac{\partial h_s}{\partial n} + q_w^{*}
&= - D_f \frac{\partial h_f}{\partial n},
\end{aligned},
\right.
\qquad \text{on } \partial \Omega_{fs}
\end{equation}
where the equivalent flux jump $\mathit{q_w^{*}}$ still exists and is given by
\begin{equation}
q_w^{*}
=
D_s \left( 1 - \frac{1}{\alpha_s} \right) \frac{\partial h_s}{\partial n}
+ q_w
=
D \left( 1 - \frac{1}{\alpha} \right) \frac{\partial h}{\partial n}
+ q_w .
\end{equation}
Then, a unified expression for the equivalent flux jump can be written as
\begin{equation}
q_w^{*}
=
D \left( 1 - \frac{1}{\alpha} \right) \frac{\partial h}{\partial n}
+ q_w ,
\end{equation}
where the equivalent coefficient is set as
\begin{equation}
\alpha(\phi_0)
=
\begin{cases}
\alpha_f = 1, & \phi_0 < 0, \\
\alpha_s, & \phi_0 \ge 0 .
\end{cases}
\label{eq19}
\end{equation}

Following our new VPM for a Neumann boundary condition introduced in Sec. \ref{subsec2_3}, the unified governing equation for conjugate scalar transport with an interfacial scalar flux jump boundary condition can be formulated as:
\begin{equation}
\frac{\partial h}{\partial t}
+ \boldsymbol{u} \cdot \nabla h
=
\nabla \cdot ( D \nabla h )
+ \nabla \cdot \bigl( q_w^{*} \, \phi \, \boldsymbol{n} \bigr) .
\end{equation}
The value of the molecular diffusivity is determined whether a point of interest lies in the fluid or solid region, which can be judged by the sign of the local values of the level-set function, 
\begin{equation}
D(\phi_0)
=
\begin{cases}
D_f, & \phi_0 < 0, \\
D_s, & \phi_0 \ge 0 .
\end{cases}
\label{eq21}
\end{equation}
The scheme is implemented in open-source codes OpenFOAM (version 8) based on a finite volume method. The second-order linear scheme is adopted for the discretization of advection, diffusion, and gradient terms. The Euler implicit scheme is employed for transient terms.

\subsection{Error analysis}
\label{subsec2_5}
\begin{figure}[H]
\centering
\includegraphics[width=0.6\textwidth]{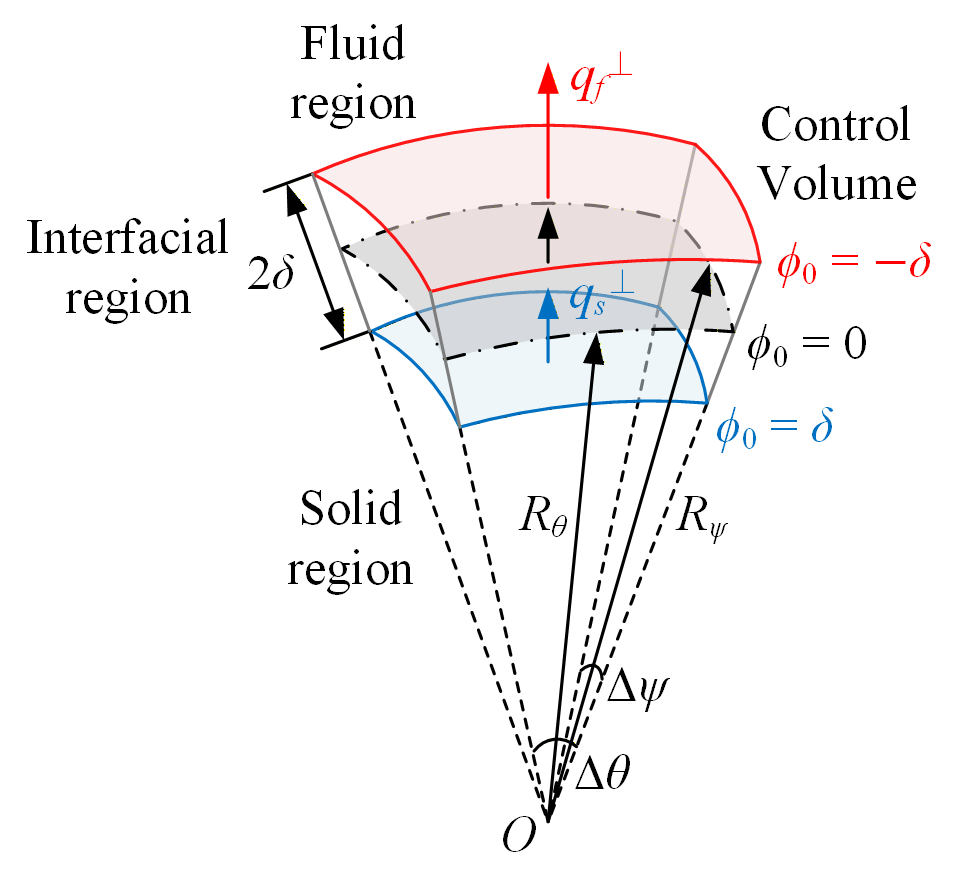}
\caption{Schematic of the control volume over the interfacial region. Here, $\psi$ and $\theta$ denote the azimuthal and the polar directions with the corresponding radius of curvature $\mathit{R_\psi}$ and $\mathit{R_\theta}$, respectively, in a three-dimensional polar coordinate system with the origin $\mathit{O}$.}\label{fig4}
\end{figure}

Even though the jump condition Eq. (\ref{eq14}) is satisfied for a flat interface as discussed in Sec. \ref{subsec2_3}, there exists another type of error due to the interfacial curvature. To analyze this, we consider a finite thickness interface whose two principal radii of curvatures are respectively $\mathit{R_\psi}$ and $\mathit{R_\theta}$ in the azimuthal $\psi$ and the polar $\theta$  directions, as illustrated in Fig. \ref{fig4}. The net scalar flux across the boundaries of the control volume (C.V.) in the wall-normal direction can be derived as
\begin{align}
\iint_{\partial \mathrm{C.V.}} q^{\perp}\, dS
&=
q_f^{\perp} (R_{\psi}+\delta)(R_{\theta}+\delta)\,\Delta\psi\,\Delta\theta
-
q_s^{\perp} (R_{\psi}-\delta)(R_{\theta}-\delta)\,\Delta\theta\,\Delta\psi
\notag
\\[6pt]
&=
(q_f^{\perp}-q_s^{\perp})(R_{\psi}R_{\theta}+\delta^{2})\,\Delta\theta\,\Delta\psi
+
(q_f^{\perp}+q_s^{\perp})(R_{\psi}+R_{\theta})\,\delta\,\Delta\theta\,\Delta\psi .
\label{eq22}
\end{align}
As the interfacial half-width $\delta$ reduces to zero, Eq. (22) will converge to zero when there is not a scalar flux jump, i.e., $\mathit{q_f^{\perp} = q_s^{\perp}}$. When there exists a scalar flux jump, the first term does not vanish, and it should balance with the VPM source term. The volume integration of the VPM term within the C.V. can be calculated as
\begin{align}
\iiint_{\mathrm{C.V.}}
\nabla \cdot \bigl( q_w \, \phi \, \boldsymbol{n} \bigr)\, dV
&=
\frac{q_w}{2\delta}
\int_{-\delta}^{\delta}
\Bigl[
(R_{\psi}+\delta)(R_{\theta}+\delta)
-
(R_{\psi}-\delta)(R_{\theta}-\delta)
\Bigr]
\,\Delta\psi\,\Delta\theta\, d\phi_0
\notag
\\[6pt]
&=
q_w
\bigl(
R_{\psi} R_{\theta}
+ \tfrac{1}{3}\delta^{2}
\bigr)
\,\Delta\theta\,\Delta\psi .
\label{eq23}
\end{align}

The following conservation between the volume integration of the VPM term in Eq. (\ref{eq23}) and the scalar flux across the boundary in Eq. (\ref{eq22}) should be satisfied within the control volume, so that
\begin{align}
0
&=
\iiint_{\mathrm{C.V.}}
\nabla \cdot \bigl( q_w \, \phi \, \boldsymbol{n} \bigr)\, dV
-
\iint_{\partial \mathrm{C.V.}} q^{\perp}\, dS
\notag
\\[6pt]
&=
\bigl( q_w - (q_f^{\perp}-q_s^{\perp}) \bigr)
\left( R_{\psi}R_{\theta} + \tfrac{1}{3}\delta^{2} \right)
\,\Delta\theta\,\Delta\psi
\notag
\\[6pt]
&\quad
-
(q_f^{\perp}-q_s^{\perp})\,\tfrac{2}{3}\delta^{2}\,\Delta\theta\,\Delta\psi
-
(q_f^{\perp}+q_s^{\perp})(R_{\psi}+R_{\theta})\,\delta\,\Delta\theta\,\Delta\psi .
\label{eq24}
\end{align}
The error between the scalar flux difference between the fluid and the solid sides and the corresponding scalar flux jump can be calculated as
\begin{equation}
q_w - \bigl( q_f^{\perp} - q_s^{\perp} \bigr)
=
\frac{(q_f^{\perp}+q_s^{\perp})(R_{\psi}+R_{\theta})}{R_{\psi}R_{\theta}+\tfrac{1}{3}\delta^{2}}\,\delta
+
\frac{\tfrac{2}{3}(q_f^{\perp}-q_s^{\perp})}{R_{\psi}R_{\theta}+\tfrac{1}{3}\delta^{2}}\,\delta^{2} .
\label{eq25}
\end{equation}
The first term in Eq. (\ref{eq25}) indicates that the jump condition is satisfied in the first-order accuracy with respect to the interfacial half-width $\delta$. Considering that $\delta$ is set as $\mathit{K_\delta} \Delta$, where $\Delta$ denotes local diagonal grid spacing, it further suggests that the error of the VPM source term is of first-order accuracy with respect to the grid size. Eq. (\ref{eq25}) also suggests that the RHS becomes null for a flat interface as $\mathit{R_\psi}$ and $\mathit{R_\theta}$ becomes infinity. As presented by Brown-Dymkoski et al \cite{Brown14}, conventional VPMs without jump conditions have the first-order accuracy, and the above analysis shows that the same order of accuracy can be achieved even with interfacial jump conditions.

\section{Verification}
\label{sec3}
In this section, a one-dimensional diffusion problem is investigated to verify the accuracy of the present VPM for the Neumann boundary condition.
\subsection{One-dimensional diffusion problem}
\label{subsec3_1}
\begin{figure}[H]
\centering
\includegraphics[width=0.8\textwidth]{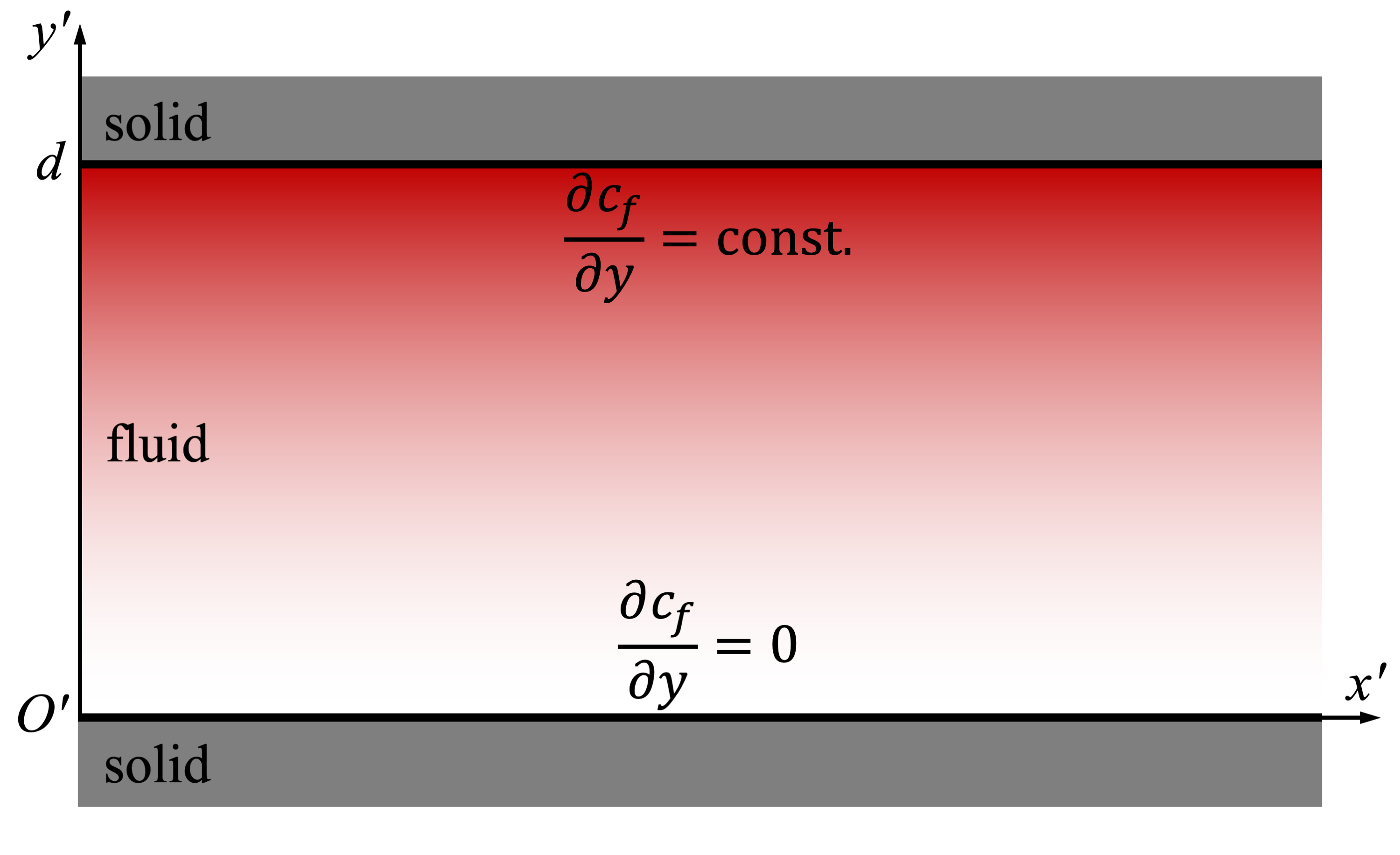}
\caption{Schematic of the one-dimensional diffusion problem.}\label{fig5}
\end{figure}

The transient one-dimensional diffusion problem in a single fluid phase displayed in Fig. \ref{fig5} is considered here. A zero-scalar-flux condition is imposed on the bottom wall, whereas a constant scalar flux of 2$\mathit{d}$ is imposed on the top wall. Here, the half-channel width $\mathit{d}$/2 is taken as the characteristic length for non-dimensionalization.

\begin{figure}[H]
\centering
\includegraphics[width=0.6\textwidth]{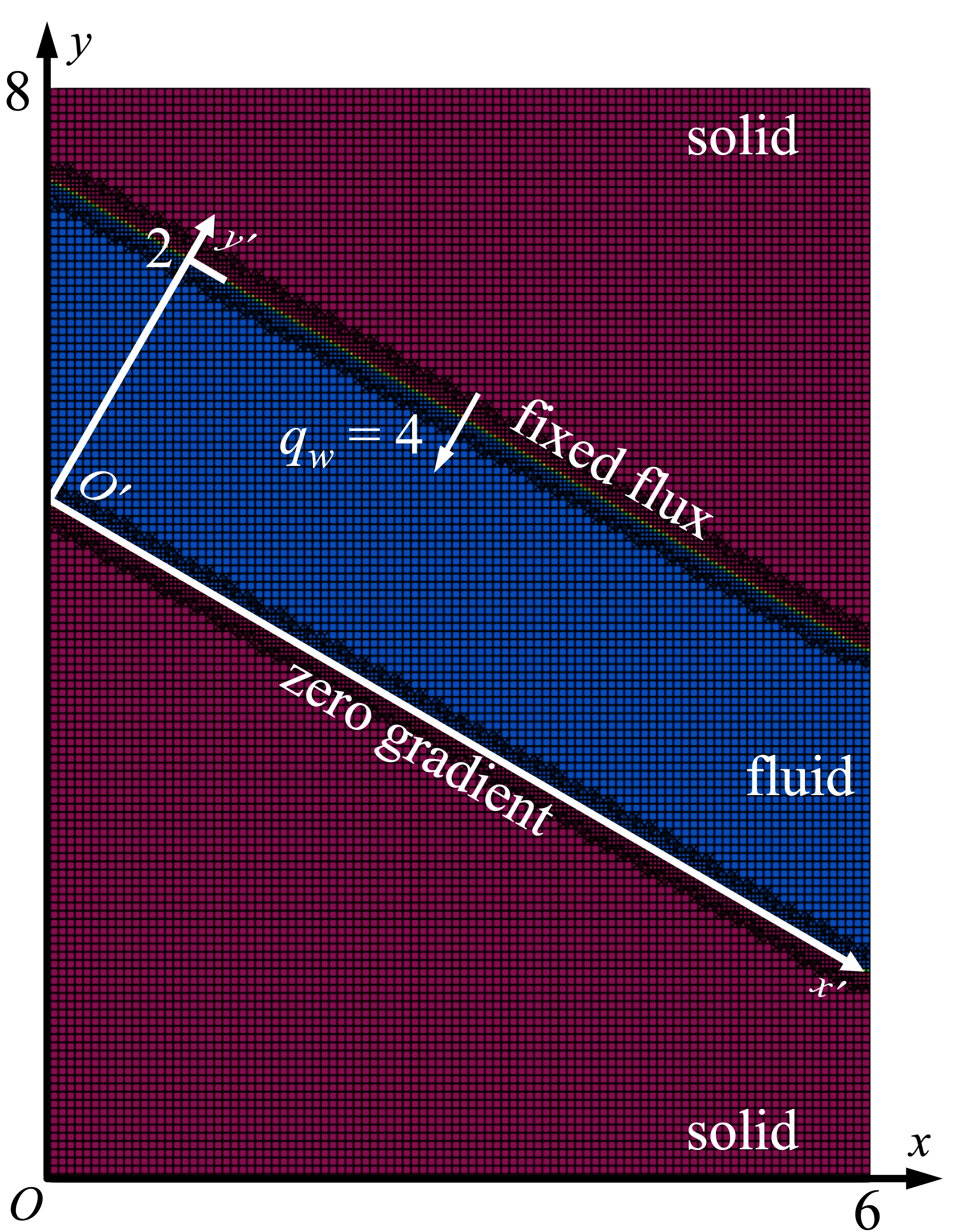}
\caption{Computational domain for the one-dimensional diffusion problem with Cartesian grids in an inclined fluid domain.}\label{fig6}
\end{figure}

\noindent
A two-dimensional computational domain with a size of [0, 6]$\times$[0, 8] is built to simulate the one-dimensional diffusion problem, and then the Cartesian grids are generated. To show the adaptability of the VPM for an arbitrary interface not aligning to the Cartesian grids, the fluid region is inclined at an angle of $30^{\circ}$ as shown in Fig. \ref{fig6}. Here, $\mathit{x}$  and $\mathit{y}$ denote the coordinates of the computational domain, while $\mathit{x'}$ and $\mathit{y'}$  correspond to those aligned in the wall tangential and normal directions to the top and bottom walls. In order to capture a rapid change of the scalar fields near the solid boundary, Automatic Mesh Refinement (AMR) based on a local value of the level-set function is employed so as to refine the grid resolution around the fluid-solid interface.

This one-dimensional diffusion problem in a single fluid phase can be described by the following mathematical models:
\begin{equation}
\frac{\partial c_f(t,y')}{\partial t}
=
D_f \frac{\partial^{2} c_f(t,y')}{\partial {y'}^{2}},
\qquad 0 \le y' \le d,\ \ t \ge 0 ,
\label{eq26}
\end{equation}
with the boundary conditions on the bottom and top walls,
\begin{equation}
\left. \frac{\partial c_f(t,y')}{\partial y'} \right|_{y'=0} = 0,
\label{eq27}
\end{equation}
\begin{equation}
\left. \frac{\partial c_f(t,y')}{\partial y'} \right|_{y'=d} = 2d.
\label{eq28}
\end{equation}
The scalar flux on the top wall is defined as twice the value of the channel width, just for the simplicity of the analytical solution of the scalar profile. The initial condition of the scalar is set to be the following parabolic distribution:
\begin{equation}
c_f(0,y') = {y'}^{2}.
\label{eq29}
\end{equation}
According to the governing equation (\ref{eq26}) and corresponding boundary (\ref{eq27}-\ref{eq28}) and initial conditions (\ref{eq29}), the analytical solution for the scalar distribution is given by
\begin{equation}
c_{\mathrm{ANL}}(t,y') = 2 D_f t + {y'}^{2} .
\label{eq30}
\end{equation}
Following the present VPM proposed in Sec. \ref{subsec2_3}, the wall scalar flux is incorporated in the governing equation as an additional source term as
\begin{equation}
\frac{\partial c}{\partial t}
=
D \frac{\partial^{2} c}{\partial y^{2}}
+ \nabla \cdot \bigl( q_w \, \phi \, \boldsymbol{n} \bigr) ,
\label{eq31}
\end{equation}
where the jump in the scalar flux on the top wall $\mathit{y' = d}$ is set as $\mathit{q_w = 2D_f d}$ according to the boundary condition Eq. (\ref{eq28}). The phase-dependent diffusivity is defined as:
\begin{equation}
D =
\begin{cases}
0, & y' < 0 \ \text{or}\ y' > d, \\
D_f, & 0 \le y' \le d .
\end{cases}
\end{equation}
By setting it to zero inside the solid, we can ensure that the scalar is only diffused into the fluid region.

\subsection{Grid convergence}
\label{subsec3_2}
\begin{figure}[H]
\centering
\includegraphics[width=0.6\textwidth]{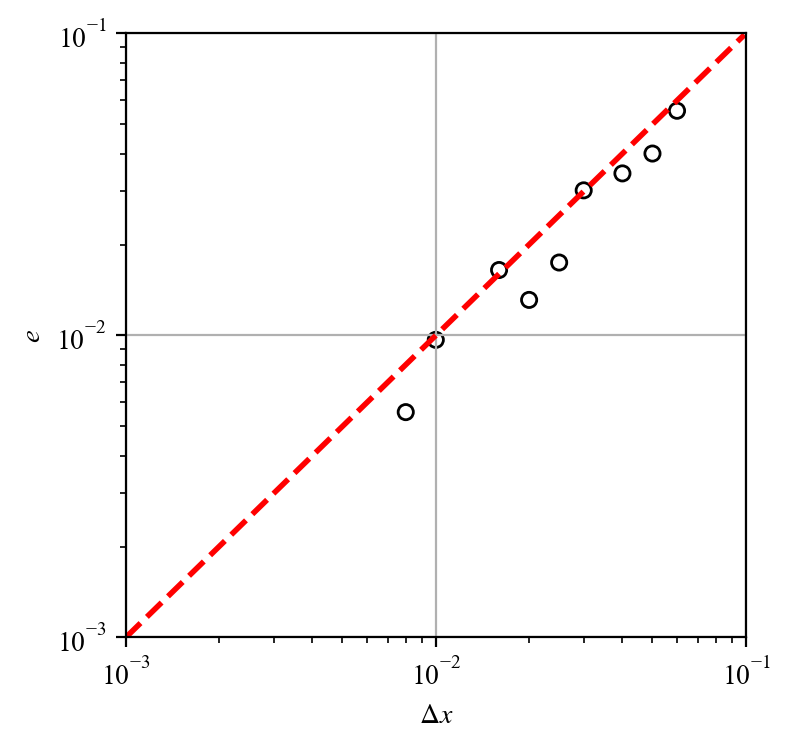}
\caption{Relative error under various grids..}\label{fig7}
\end{figure}

A systematic study on grid convergence is performed with the grid size $\Delta x$ gradually reducing from 0.06 to 0.008. The simulation results by the present VPM are compared with the analytical solutions to check the accuracy, and the following relative error is defined,
\begin{equation}
e
=
\int_{0}^{1}
\int_{0}^{d}
\left|
\frac{c^{\mathrm{VPM}}(t,y') - c^{\mathrm{ANL}}(t,y')}
     {c^{\mathrm{ANL}}(t,y')}
\right|
\, dy'\, dt,
\end{equation}
where the superscripts of VPM and ANL denote results by the present simulations and the analytical solutions (\ref{eq30}), respectively. Here, since two-dimensional simulations are performed by means of VPM, the result of $c^{\mathrm{VPM}}(t,y')$ is calculated as a function of $\mathit{y'}$ by taking the average along $\mathit{x'}$ direction as
\begin{equation}
c^{\mathrm{VPM}}(t,y')
=
\frac{1}{L_{x'}}
\int_{l}
c^{\mathrm{VPM}}(t,x',y')\, dx' .
\end{equation}
Figure \ref{fig7} summarizes the relative errors of the present VPM for different grid sizes. The red dashed line in Fig. \ref{fig7} suggests the first-order accuracy, i.e., $\mathit{e \propto \mathrm{\Delta} x}$, which is the same as the existing VPM \cite{Brown14}.

\subsection{Simulation results}
\label{subsec3_3}
\begin{figure}[H]
\centering

\makebox[\linewidth]{%
\raisebox{0.5\height}{\textbf{(a)}}\hspace{1em}%
\includegraphics[width=0.45\linewidth]{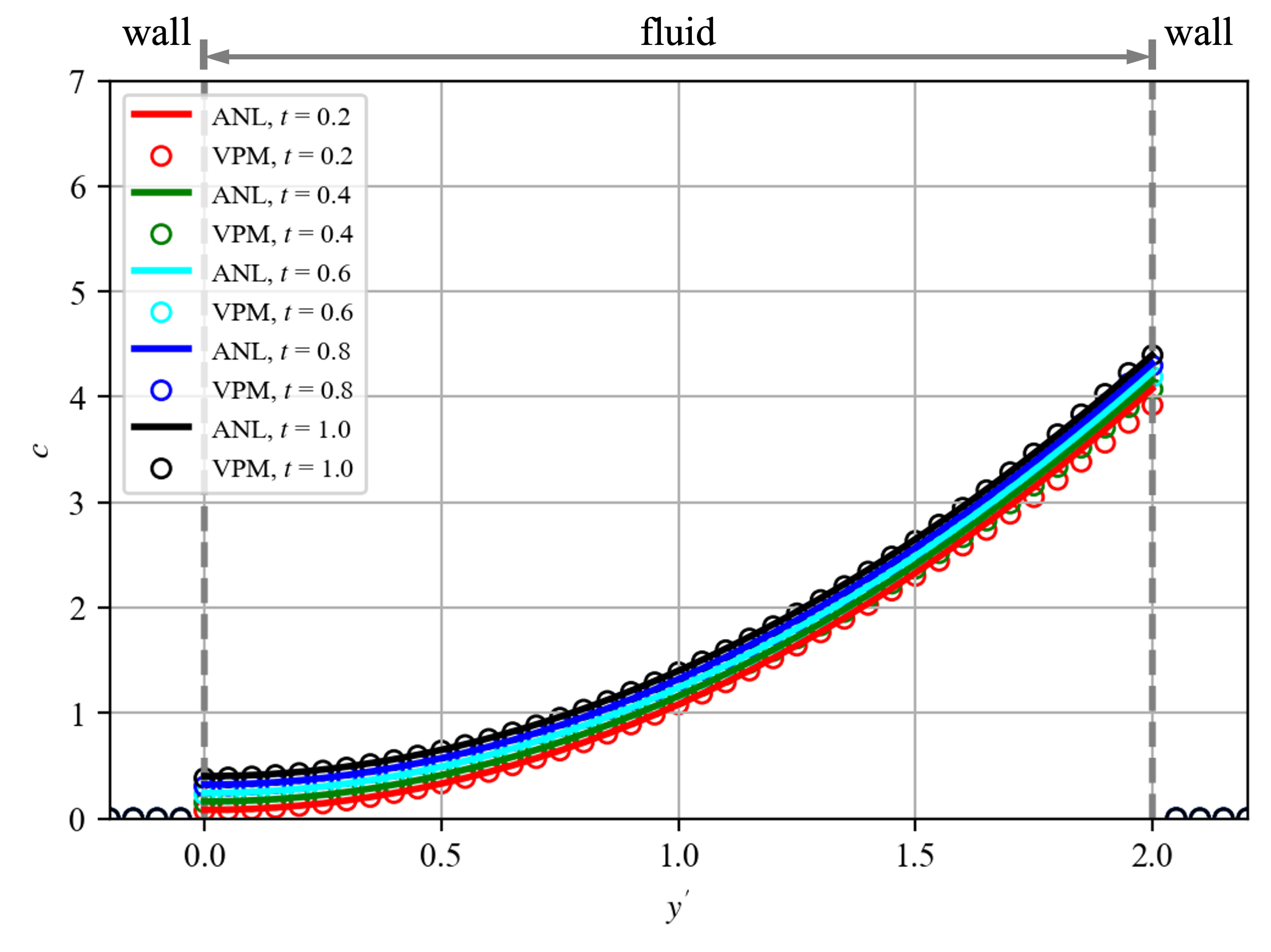}%
\hfill
\includegraphics[width=0.45\linewidth]{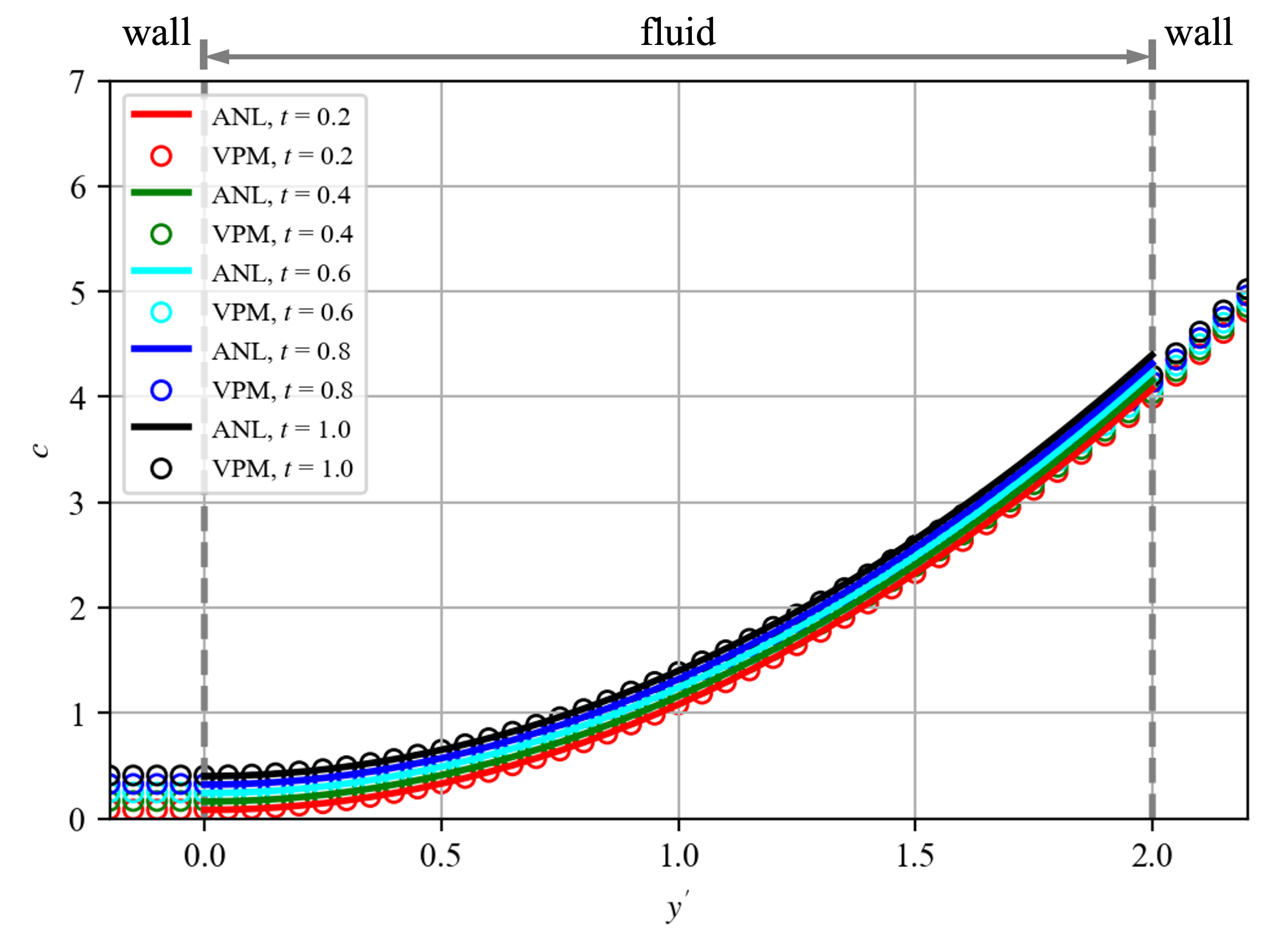}%
}

\vspace{1ex}

\makebox[\linewidth]{%
\raisebox{0.5\height}{\textbf{(b)}}\hspace{1em}%
\includegraphics[width=0.45\linewidth]{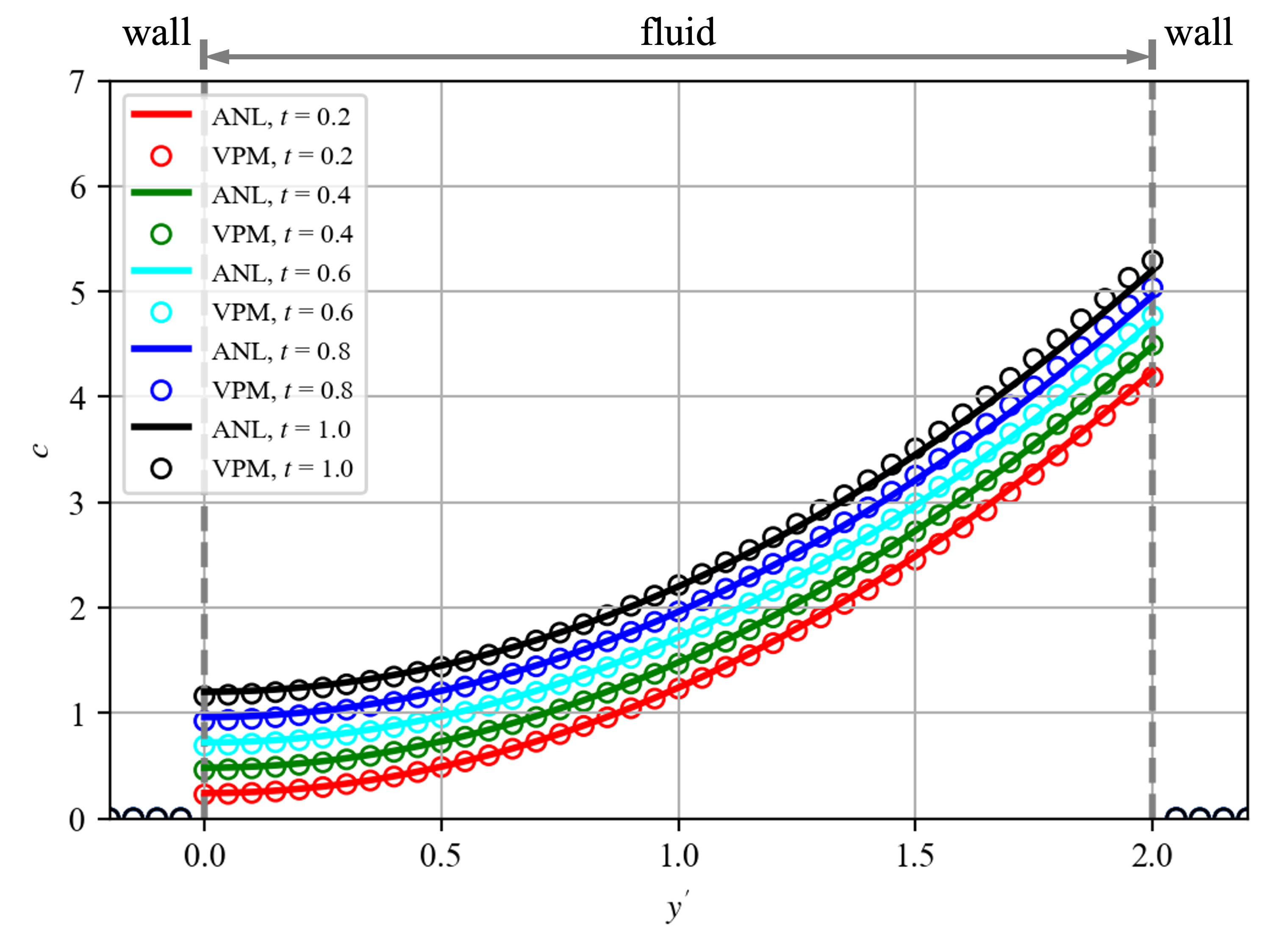}%
\hfill
\includegraphics[width=0.45\linewidth]{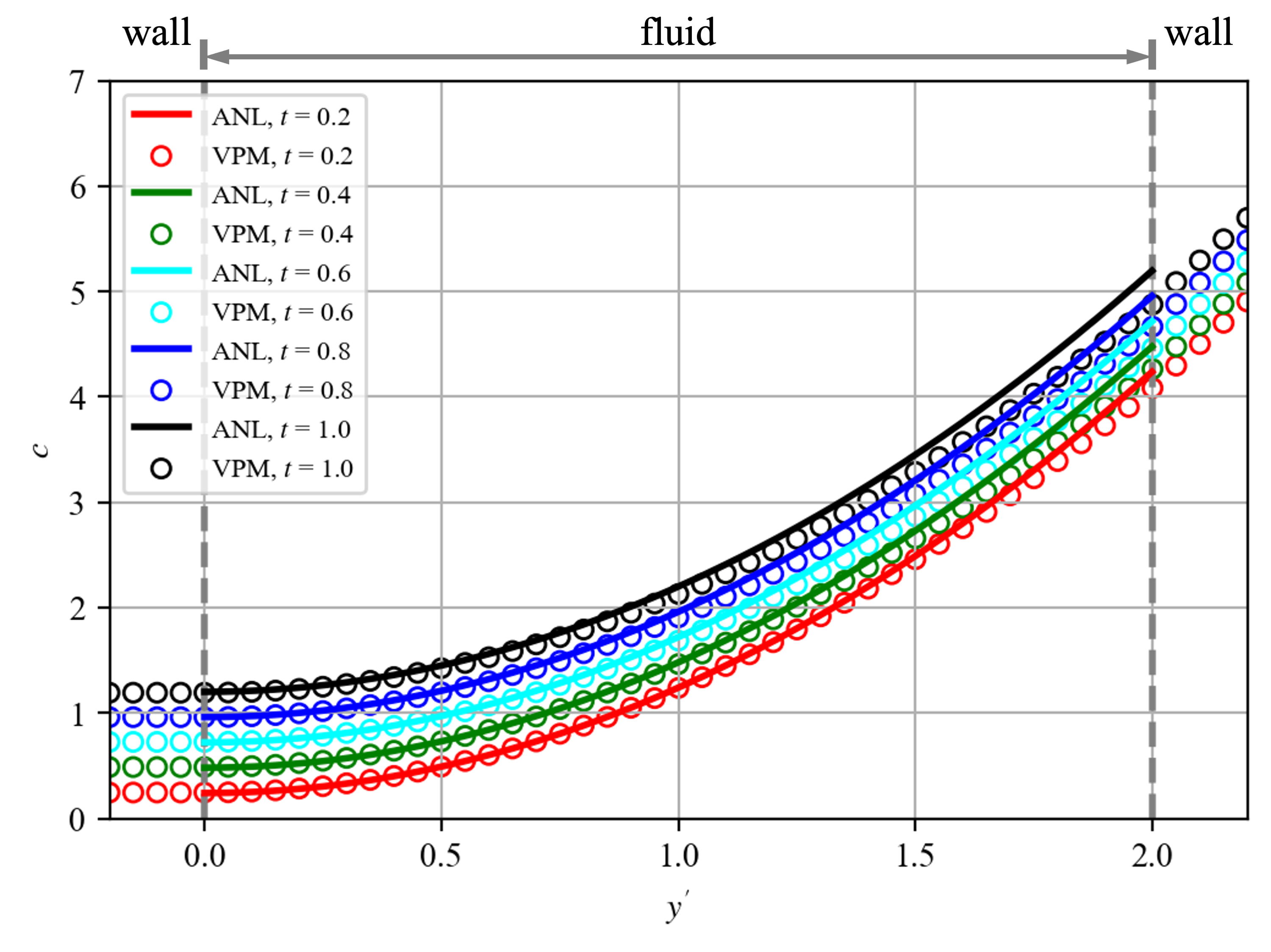}%
}

\vspace{1ex}

\makebox[\linewidth]{%
\raisebox{0.5\height}{\textbf{(c)}}\hspace{1em}%
\includegraphics[width=0.45\linewidth]{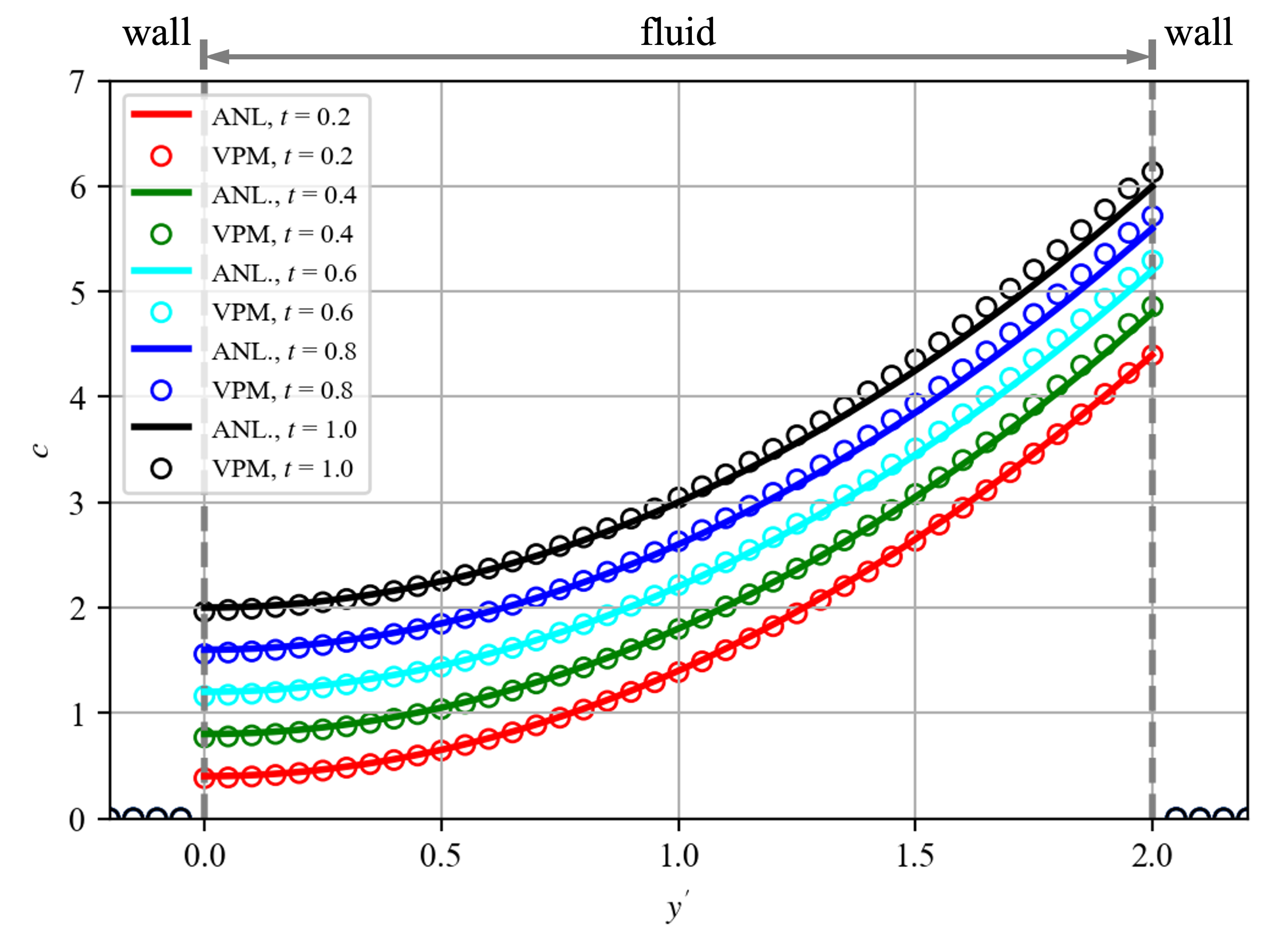}%
\hfill
\includegraphics[width=0.45\linewidth]{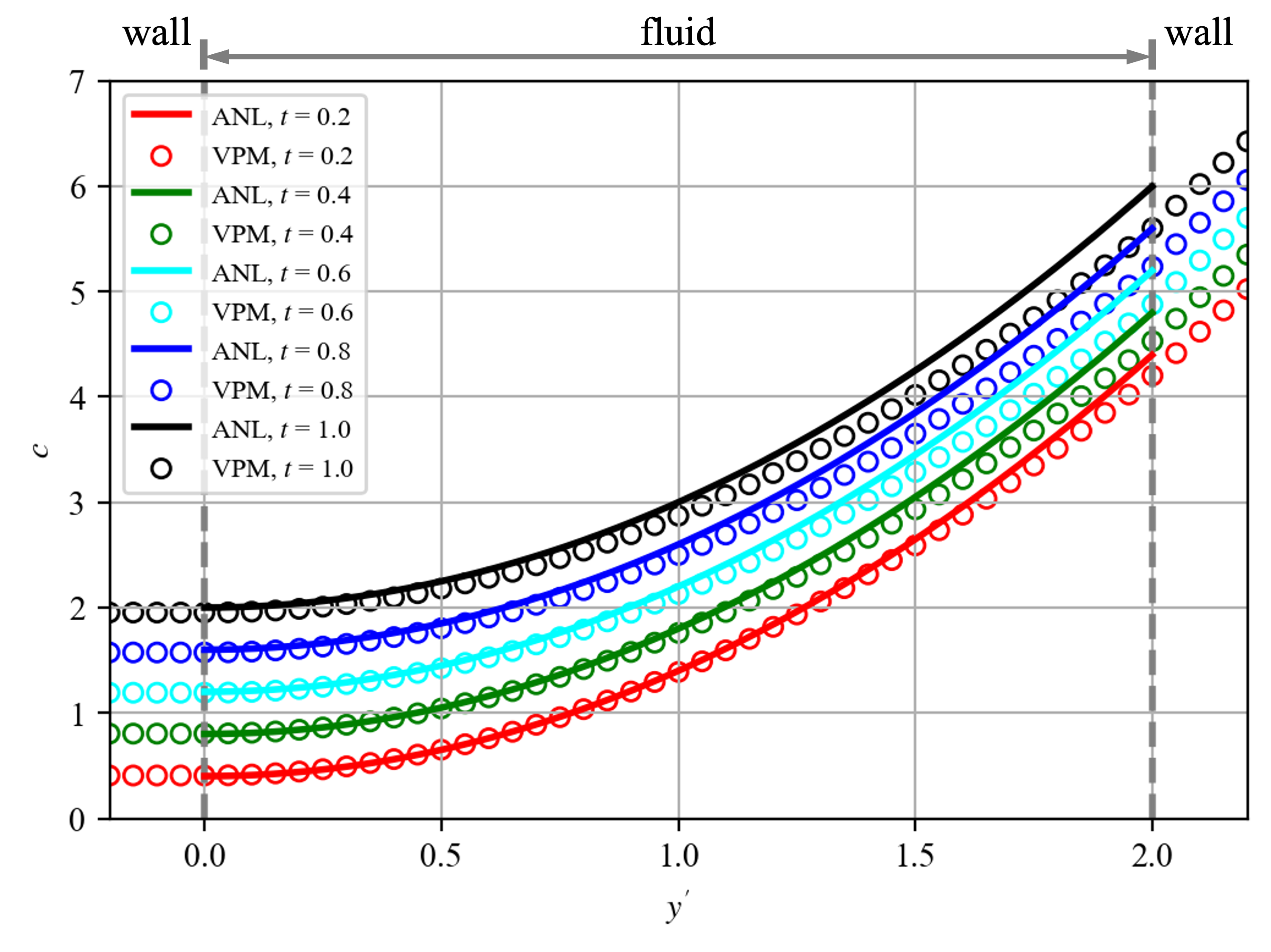}%
}

\caption{Simulation results by the present VPM for various fluid molecular diffusivity of $\mathit{D_f}$ = (a) 0.2, (b) 0.6, and (c) 1.0. Here, the left and right columns plot the results using the present and existing VPMs, respectively.}
\label{fig8}
\end{figure}

Based on the results of the grid convergence study in Sec. \ref{subsec3_2}, the grid size of $\Delta x$  = 0.02 is adopted in the following analyses, considering a balance between calculation cost and accuracy. Figure 8 shows the comparison between the results of the present VPM and the analytical solutions for different values of the fluid molecular diffusivity $\mathit{D_f}$. Since the transient problem is considered here, the time evolutions of the scalar distribution for the initial time period from $\mathit{t}$ = 0.2 to $\mathit{t}$ = 1.0 are plotted. Good agreement between the simulation results and the analytical solutions under various diffusivity can generally be confirmed, as shown in the left column of Fig. 8. Further quantitative analysis shows that the relative errors for different values of molecular diffusivity $\mathit{D_f}$ = 0.2, 0.4, 0.6, 0.8, 1.0 are 1.22\%, 1.17\%, 1.23\%, 1.29\%, and 1.34\%, respectively, with their mean value of 1.25\%. The existing VPM in Ref.\cite{Brown14} is also implemented in OpenFOAM, and simulations are performed under the same conditions using the same mesh. The simulation results are shown in the right column of Fig. \ref{fig8}. The relative errors by the present and existing VPMs are calculated and summarized in Table \ref{tab1}.

\begin{table}[H]
\centering
\caption{Relative errors by the present and VPMs.}
\label{tab1}
\resizebox{\linewidth}{!}{%
\begin{tabular}{l c c c c c c}
\toprule
 & $D_f=0.2$ & $D_f=0.4$ & $D_f=0.6$ & $D_f=0.8$ & $D_f=1.0$ & \textbf{Mean} \\
\midrule
Present Scheme  & 1.22\% & 1.17\% & 1.23\% & 1.29\% & 1.34\% & \textbf{1.25\%} \\
Existing Scheme & 1.02\% & 1.64\% & 2.20\% & 2.68\% & 3.08\% & \textbf{2.12\%} \\
\bottomrule
\end{tabular}
}
\end{table}

From the right column in Fig. \ref{fig8}, it can be seen that scalar distribution appears even inside the solid regions ($\mathit{y'}$  < 0 and $\mathit{y'}$  > 2) when the existing VPM is used, while it is kept exactly zero in the present scheme as shown in left column of Fig. 8. This is because the VPM source term in the existing scheme has non-zero values not only in the interfacial region but also in the solid region. As indicated in Table 1, for the present scheme, the relative error of the present scheme remains almost at the same level for different values of diffusivity. In contrast, the relative error of the existing scheme gradually increases as the molecular diffusivity rises. This would be attributed to the large extension of the non-physical distribution in the solid region with increasing diffusivity. The generation of the non-physical distribution inside the solid regions is not suitable when fluid-solid coupled problems are considered, as will be shown later.

\section{Fluid-solid coupled diffusion problem}
\label{sec4}
In this section, the VPM for the Neumann boundary condition proposed in Sec. 3 is further extended to fluid-solid coupled problems with interfacial jump conditions.
\subsection{Problem description}
\label{subsec4_1}
\begin{figure}[H]
\centering
\includegraphics[width=0.8\textwidth]{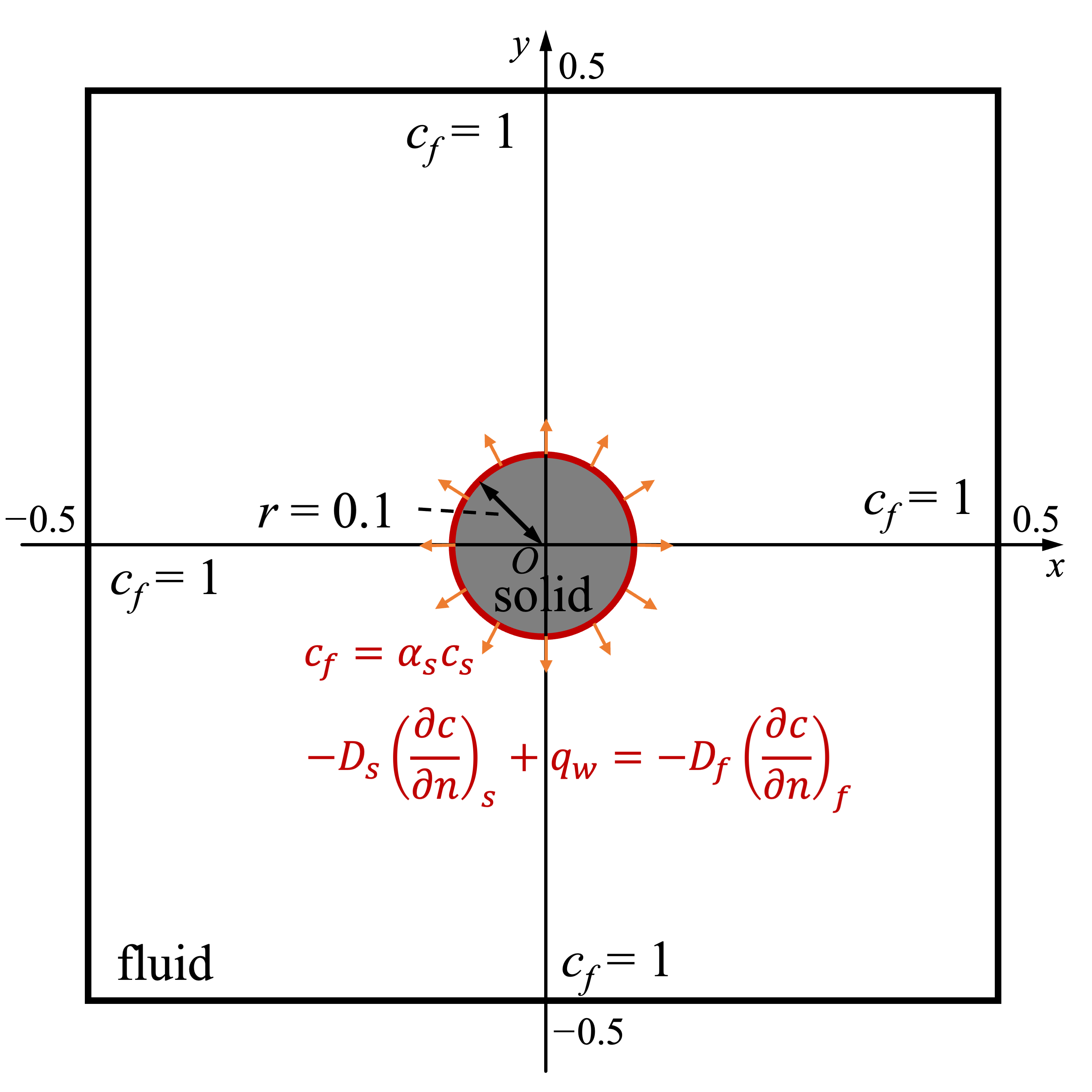}
\caption{Schematic of the fluid-solid coupled diffusion problem. Here, the white and grey regions represent fluid and solid phases, respectively, and the red circle denotes the fluid-solid interface. The orange arrows show the uniform distribution of constant scalar flux $\mathit{q_w}$ on the interface.}\label{fig9}
\end{figure}

A two-dimensional transient conjugate transport problem for scalar $\mathit{c}$ that diffuses in coupled fluid-solid regions with jumps in the interfacial scalar and its flux is considered here, as illustrated in Fig. \ref{fig9}. A solid circular cylinder with a radius of $\mathit{r}$  = 0.1 is located at the center of a square computational domain with the unit size of [-0.5, 0.5]$\times$[-0.5, 0.5]. The origin of the coordinates is defined at the center of the cylinder. A constant scalar flux jump $\mathit{q_w}$ is assumed at the interface. This may correspond to a situation where a scalar is generated or dissipated constantly at the interface due to a chemical reaction or phase change. The discontinuity of the scalar at the interface with a multiplication factor $\alpha_s$ is also considered here. Zero-scalar-flux conditions are imposed on the four boundaries of the computational domain. This problem can be formulated as follows:
\begin{equation}
\left\{
\begin{aligned}
\frac{\partial c_f}{\partial t}
&= \nabla \cdot \bigl( D_f \nabla c_f \bigr),
\qquad \text{in } \Omega_f, \\
\frac{\partial c_s}{\partial t}
&= \nabla \cdot \bigl( D_s \nabla c_s \bigr),
\qquad \text{in } \Omega_s ,
\end{aligned}
\right.
\end{equation}
under the boundary conditions of:
\begin{equation}
\left\{
\begin{aligned}
c_f &= \alpha_s c_s, \\
- D_s \left. \frac{\partial c}{\partial n} \right|_{s}
+ q_w
&=
- D_f \left. \frac{\partial c}{\partial n} \right|_{f},
&& \text{on } \partial \Omega_{fs}, \\
c_f &= 1,
&& \text{on } \partial \Omega ,
\end{aligned}
\right.
\end{equation}
and also initial conditions of:
\begin{equation}
\left\{
\begin{aligned}
c_f(t=0) &= 1, \qquad \text{in } \Omega_f, \\
c_s(t=0) &= \frac{1}{\alpha_s}, \qquad \text{in } \Omega_s .
\end{aligned}
\right.
\end{equation}
This initial condition corresponds to an equilibrium state when there exists no scalar flux jump $\mathit{q_w}$, so that the scalar field does not change if $\mathit{q_w}$ = 0. Note that, although we assume a constant scalar flux jump in the present problem, it can be easily extended to problems with a time-dependent scalar flux jump.

The unified VPM governing equation in both two phases can be written as
\begin{equation}
\frac{\partial h}{\partial t}
=
\nabla \cdot ( D \nabla h )
+ \nabla \cdot \bigl( q_w^{*} \, \phi \, \boldsymbol{n} \bigr)
\end{equation} ,
where the jump of the interfacial scalar flux is defined as follows:
\begin{equation}
q_w^{*}
=
D \left( 1 - \frac{1}{\alpha} \right) \frac{\partial h}{\partial n}
+ q_w .
\end{equation}
The phase-dependent equivalent coefficient  $\alpha$ and molecular diffusivity $\mathit{D}$ follow the same definition in Eqs. (\ref{eq19}) and (\ref{eq21}), respectively.

Totally four cases, as summarized in Table 2, are considered. Among these four cases, Cases 1 and 2 correspond to situations where only a scalar flux jump exists, i.e.,  $\alpha_s$ = 1, while the other two cases (Cases 3 and 4) consider both jumps in the scalar and its flux.

\begin{table}[H]
\centering
\caption{Fluid-solid coupled diffusion case settings.}
\label{tab2}
\setlength{\tabcolsep}{14pt}
\renewcommand{\arraystretch}{1.6}

\begin{tabular}{>{\bfseries}l c c c c c}
\toprule
 & $D_f$ & $D_s$ & $\alpha_s$ & $q_w$ & \begin{tabular}[c]{@{}c@{}}\textbf{Relative}\\ \textbf{deviation}\end{tabular} \\
\midrule
Case 1 & 0.01 & 0.05 & 1.0  & 1.0 & 2.91\% \\
Case 2 & 0.05 & 0.01 & 1.0  & 1.0 & 1.23\% \\
Case 3 & 0.01 & 0.05 & 0.8  & 1.0 & 0.85\% \\
Case 4 & 0.01 & 0.05 & 1.25 & 1.0 & 0.85\% \\
\bottomrule
\end{tabular}
\end{table}

\subsection{Convergence study on grid resolution and time step}
\label{subsec4_2}
\begin{figure}[H]
\centering
\includegraphics[width=\textwidth]{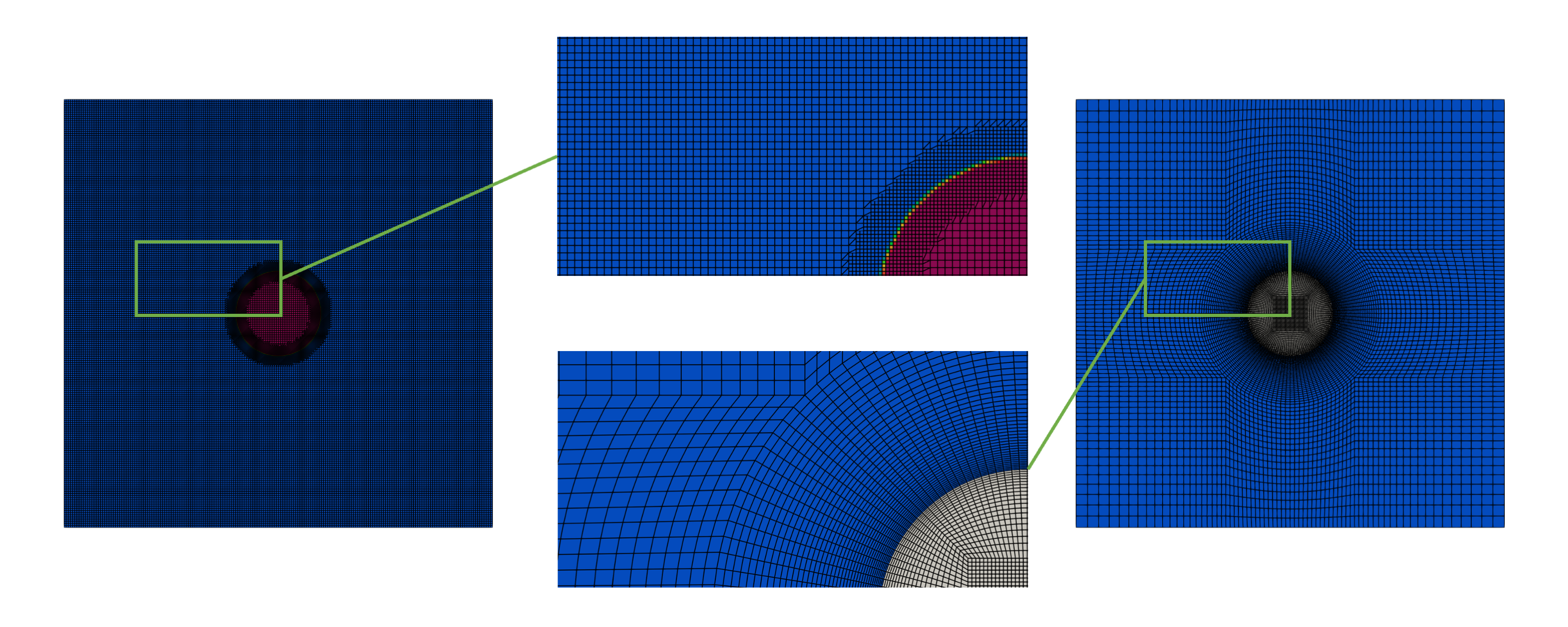}
\caption{Computational grids used in present VPM (left) and BFM (right).}\label{fig10}
\end{figure}

Cartesian grids with 1st-level AMR are employed for the present VPM. For its validation, corresponding BFM simulations are also conducted with a standard built-in OpenFOAM solver, chtMultiRegionFoam. The distributions of the Cartesian grids and body-fitted grids are shown in Fig. \ref{fig10}. The grid-convergence studies for both VPM and BFM are performed in Case 4. The results are summarized in Tables \ref{tab3} and \ref{tab4}, respectively. For VPM, the Cartesian grid resolutions along $\mathit{x}$ and $\mathit{y}$ directions are systematically increased as  $\mathit{N}_{x} \times \mathit{N}_{y}$ = 100$\times$100, 150$\times$150, 200$\times$200, 250$\times$250, and 300$\times$300. For BFM, the total number of grid points is also increased as 17,700, 30,596, 48,400, 68,336, and 93,700. For each type of grid, the simulation results using the finest one are taken as references. The relative deviation compared to the finest-grid results is defined as
\begin{equation}
\mathrm{relative\ deviation}
=
\int_{0}^{1}
\int_{-0.5}^{0.5}
\left|
\frac{c(t,x,y=0) - c^{\mathrm{finest}}(t,x,y=0)}
     {c^{\mathrm{finest}}(t,x,y=0)}
\right|
\, dx\, dt ,
\end{equation}
where the superscript "finest" refers to the simulation results obtained by the finest grids. For Cartesian grids finer than 200×200 and body-fitted grids finer than 48,400, the relative deviations are less than 0.5\%, and therefore these two meshes are adopted in the following analysis. Also, the distributions of the scalar along the line of $\mathit{y}$ = 0 for different resolutions of Cartesian and body-fitted grids are shown in Fig. 11. It is shown in Fig. 11(a) that, as we refine the Cartesian grids, the VPM results converge to the BFM ones.

\begin{table}[htbp]
\centering
\caption{Grids convergence check for VPM.}
\label{tab3}
\resizebox{\linewidth}{!}{%
\begin{tabular}{c c c c}
\toprule
$\mathit{N_x \times N_y}$ &
\makecell{\textbf{Minimum grid size}\\ \textbf{after AMR}} &
\makecell{\textbf{Number of grids}\\ \textbf{points after AMR}} &
\makecell{\textbf{Relative deviation to}\\ \textbf{the finest results}} \\
\midrule
100$\times$100 & 0.0050 & 22,368  & 1.45\% \\
150$\times$150 & 0.0033 & 48,624  & 0.67\% \\
200$\times$200 & 0.0025 & 87,128  & 0.43\% \\
250$\times$250 & 0.0020 & 131,040 & 0.31\% \\
300$\times$300 & 0.0017 & 187,200 & 0.00\% \\
\bottomrule
\end{tabular}
}
\end{table}

\begin{table}[htbp]
\centering
\caption{Grids convergence check for BFM.}
\label{tab4}
\setlength{\tabcolsep}{24pt}
\renewcommand{\arraystretch}{1.4}
\begin{tabular}{c c}
\toprule
\textbf{Number of grids points} &
\makecell{\textbf{Relative deviation to}\\ \textbf{the finest results}} \\
\midrule
17,700 & 3.74\% \\
30,596 & 2.37\% \\
48,400 & 0.47\% \\
68,336 & 0.30\% \\
93,700 & 0.00\% \\
\bottomrule
\end{tabular}
\end{table}

\begin{figure}[H]
\centering
\begin{subfigure}{0.48\linewidth}
    \caption{}
    \includegraphics[width=\linewidth]{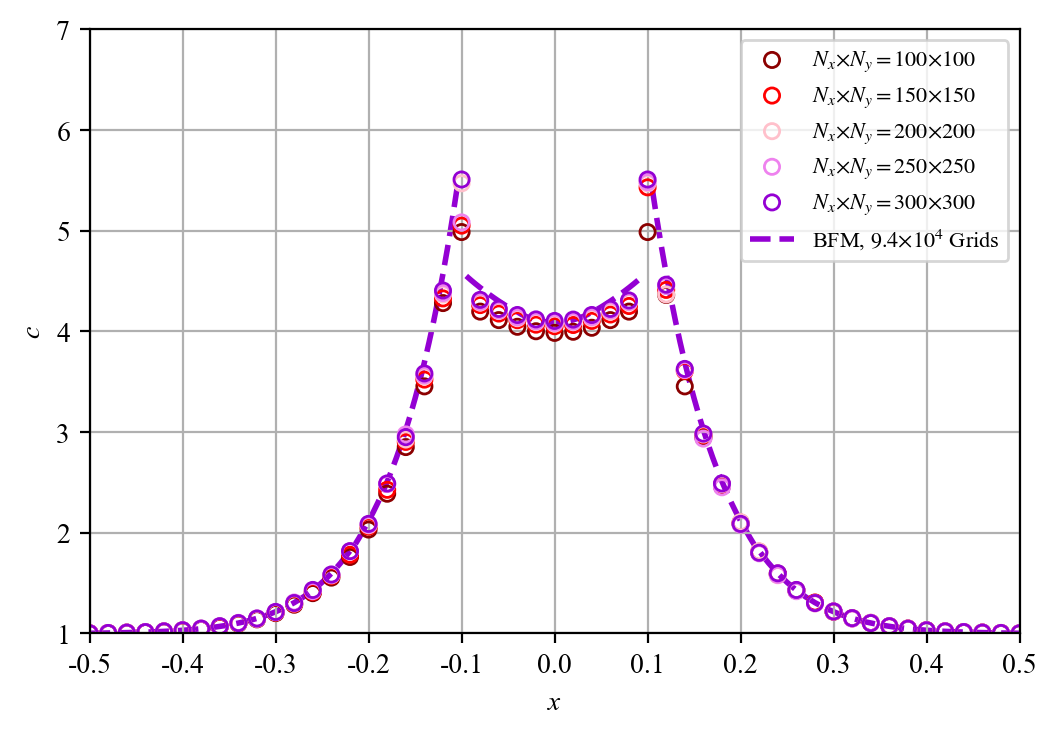}
    \centering
\end{subfigure}
\hfill
\begin{subfigure}{0.48\linewidth}
    \caption{}
    \includegraphics[width=\linewidth]{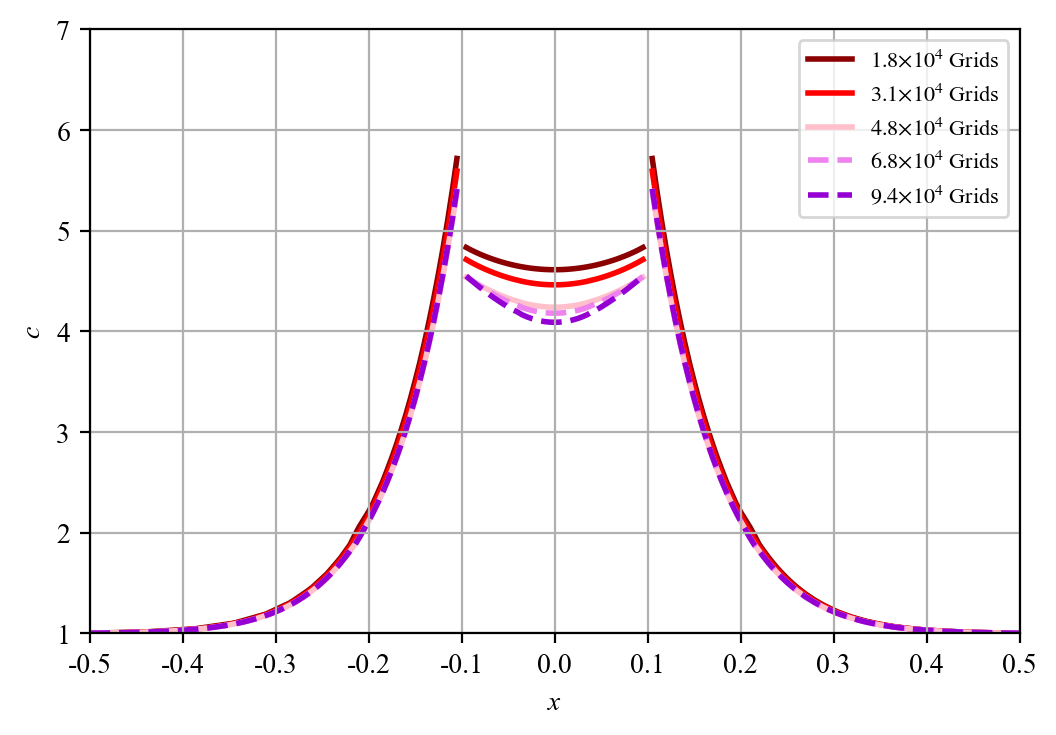}
    \centering
\end{subfigure}

\caption{Distribution of the scalar along the line $\mathit{y}$ = 0 at $\mathit{t}$ = 1.0 for different resolutions of (a) Cartesian grids and (b) body-fitted grids. In Fig. 11(a), the BFM results using the finest grids are also plotted in dashed lines as a reference.}
\label{fig11}
\end{figure}

\begin{figure}[H]
\centering
\begin{subfigure}{0.48\linewidth}
    \caption{}
    \includegraphics[width=\linewidth]{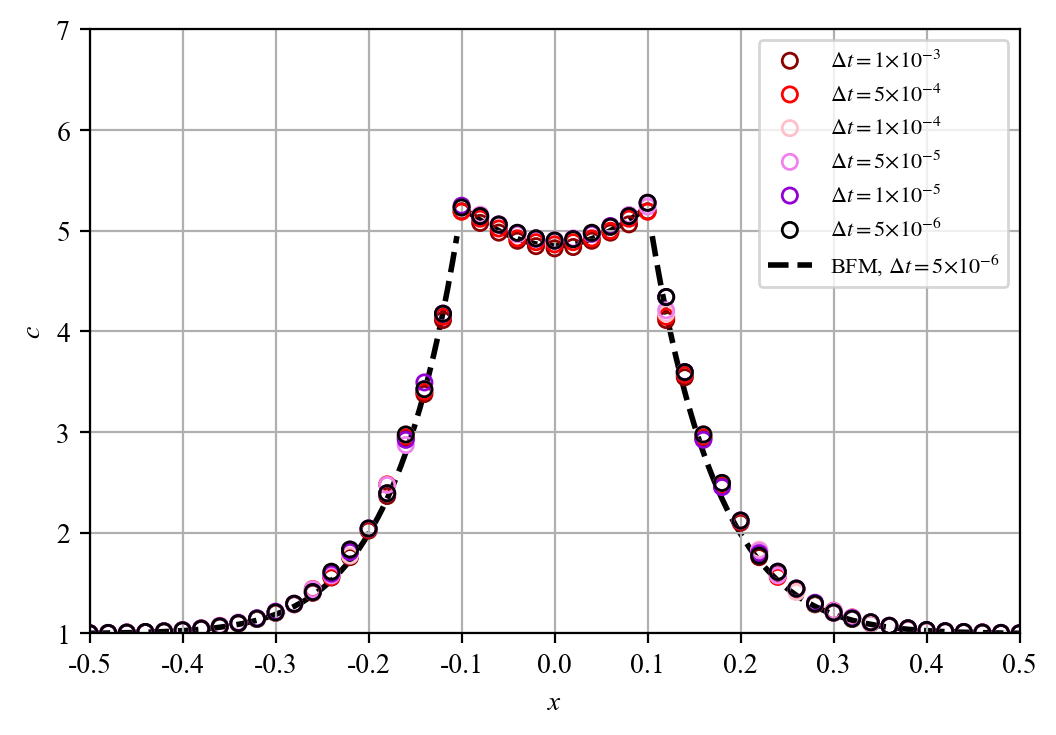}
    \centering
\end{subfigure}
\hfill
\begin{subfigure}{0.48\linewidth}
    \caption{}
    \includegraphics[width=\linewidth]{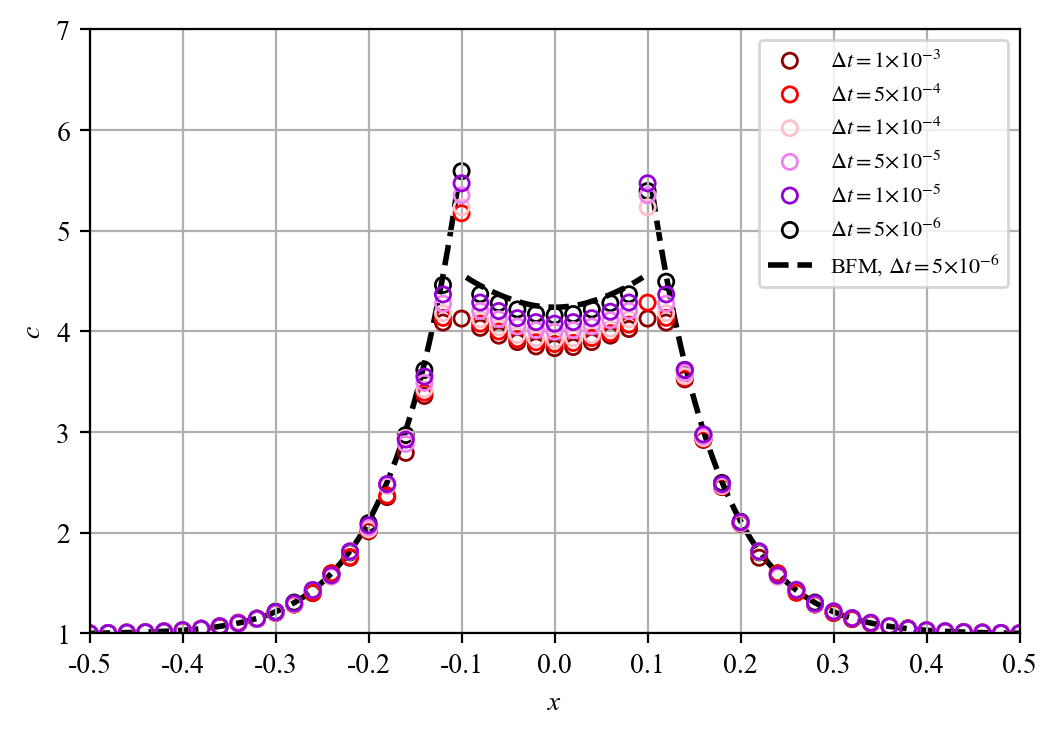}
    \centering
\end{subfigure}

\caption{Distribution of the scalar along the line $\mathit{y}$ = 0 at $\mathit{t}$ = 1.0 for (a) Case 1 and (b) Case 4 using different values of the time step. Here, the BFM results using the smallest timestep are also plotted in dashed lines as a reference.}
\label{fig12}
\end{figure}

The influence of a time step used in the present VPM is also investigated for Case 1 with scalar flux jump only and Case 4 with jumps in both scalar and its flux. The relative deviations to the results obtained by the smallest timestep are summarized in Table \ref{tab5}. Here, the relative deviation compared to the simulation results using the smallest time step is defined as,

\begin{equation}
\mathrm{relative\ deviation}
=
\int_{0}^{1}
\int_{-0.5}^{0.5}
\left|
\frac{c(t,x,y=0) - c^{\mathrm{smallest}}(t,x,y=0)}
     {c^{\mathrm{smallest}}(t,x,y=0)}
\right|
\, dx\, dt ,
\end{equation}
where the superscript "smallest" refers to the simulation results with the smallest time step. The distributions of the scalar along the line $\mathit{y}$ = 0 using different time steps are shown in Fig. \ref{fig12}. It suggests that, for the case with scalar flux jump only, the time step of $1\times10^{-4}$ is sufficient to obtain converged results, while a smaller time step of $1\times10^{-5}$ is required when both scalar and its flux jumps are considered.

\begin{table}[H]
\centering
\caption{Influence of transient timestep on simulation results.}
\label{tab5}
\resizebox{\linewidth}{!}{%
\begin{tabular}{c c c}
\toprule
 & \multicolumn{2}{c}{\textbf{Relative deviation to the results using the smallest timestep}} \\
\cmidrule(lr){2-3}
\textbf{Transient timestep} & \textbf{Case 1} & \textbf{Case 4} \\
\midrule
$1\times10^{-3}$ & 1.16\% & 3.04\% \\
$5\times10^{-4}$ & 0.82\% & 2.58\% \\
$1\times10^{-4}$ & 0.61\% & 2.18\% \\
$5\times10^{-5}$ & 0.49\% & 1.48\% \\
$1\times10^{-5}$ & 0.37\% & 0.79\% \\
$5\times10^{-6}$ & 0.00\% & 0.00\% \\
\bottomrule
\end{tabular}
}
\end{table}

\subsection{Results and discussions}
\label{subsec4_3}
The relative deviation along the line $\mathit{y}$ = 0 between VPM and BFM results defined below is shown in Table \ref{tab2}.
\begin{equation}
\mathrm{relative\ deviation}
=
\int_{0}^{1}
\int_{-0.5}^{0.5}
\left|
\frac{c^{\mathrm{VPM}}(t,x,y=0) - c^{\mathrm{BFM}}(t,x,y=0)}
     {c^{\mathrm{BFM}}(t,x,y=0)}
\right|
\, dx\, dt,
\label{eq42}
\end{equation}
where the superscripts of VPM and BFM denote simulation results by VPM and BFM, respectively. As shown in Table \ref{tab2}, the relative deviation between VPM and BFM is within 3.0\% for all four cases, with their average value of 1.46

Figure \ref{fig13} shows the time evolutions of the scalar along the line $\mathit{y}$ = 0 for all four cases. Good agreement between the present VPM results and those obtained by BFM can be found. This verifies the accuracy of the present VPM. As shown in Fig. \ref{fig13}, the largest deviation always appears at the interface ($\mathit{x}$ = -0.1, 0.1), which also affects the simulation accuracy inside the solid region (-0.1 < $\mathit{x}$ < 0.1). Despite this, the scalar distribution in the fluid region (-0.5 < $\mathit{x}$ < -0.1, and 0.1 < $\mathit{x}$ < 0.5) is produced accurately, and that in the solid region is consistent with BFM results. Since VPM is a type of diffuse-interface immersed boundary method, the accuracy within the interfacial region is not generally guaranteed. However, as we refine the mesh to reduce the thickness of the interfacial region, the results will eventually converge to the prediction by BFM, as shown in Fig. \ref{fig11}(a).

\begin{figure}[H]
\centering

\begin{subfigure}{\linewidth}
  \centering
  \caption{}
  \includegraphics[width=0.8\linewidth]{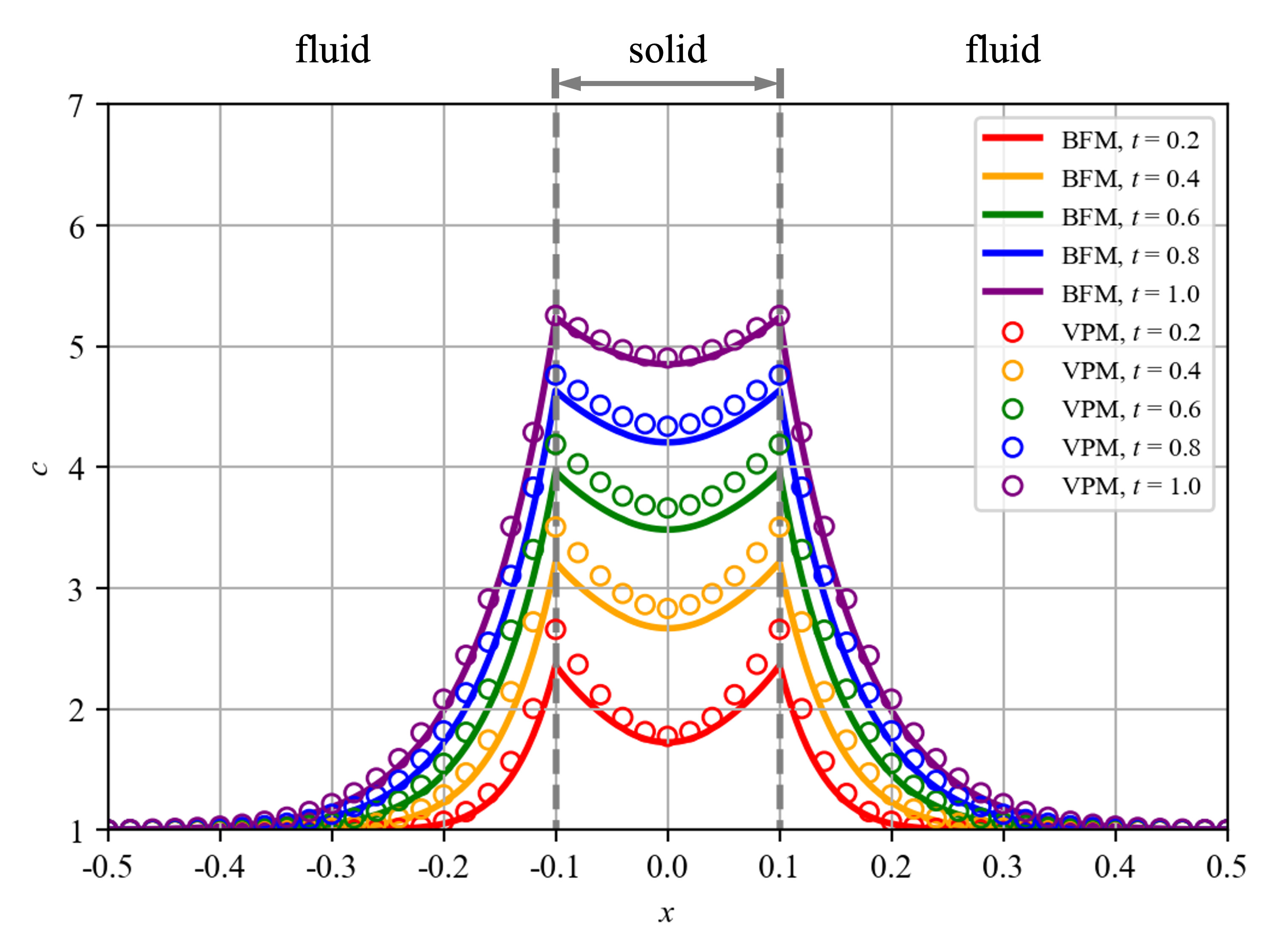}
\end{subfigure}

\medskip

\begin{subfigure}{\linewidth}
  \centering
  \caption{}
  \includegraphics[width=0.8\linewidth]{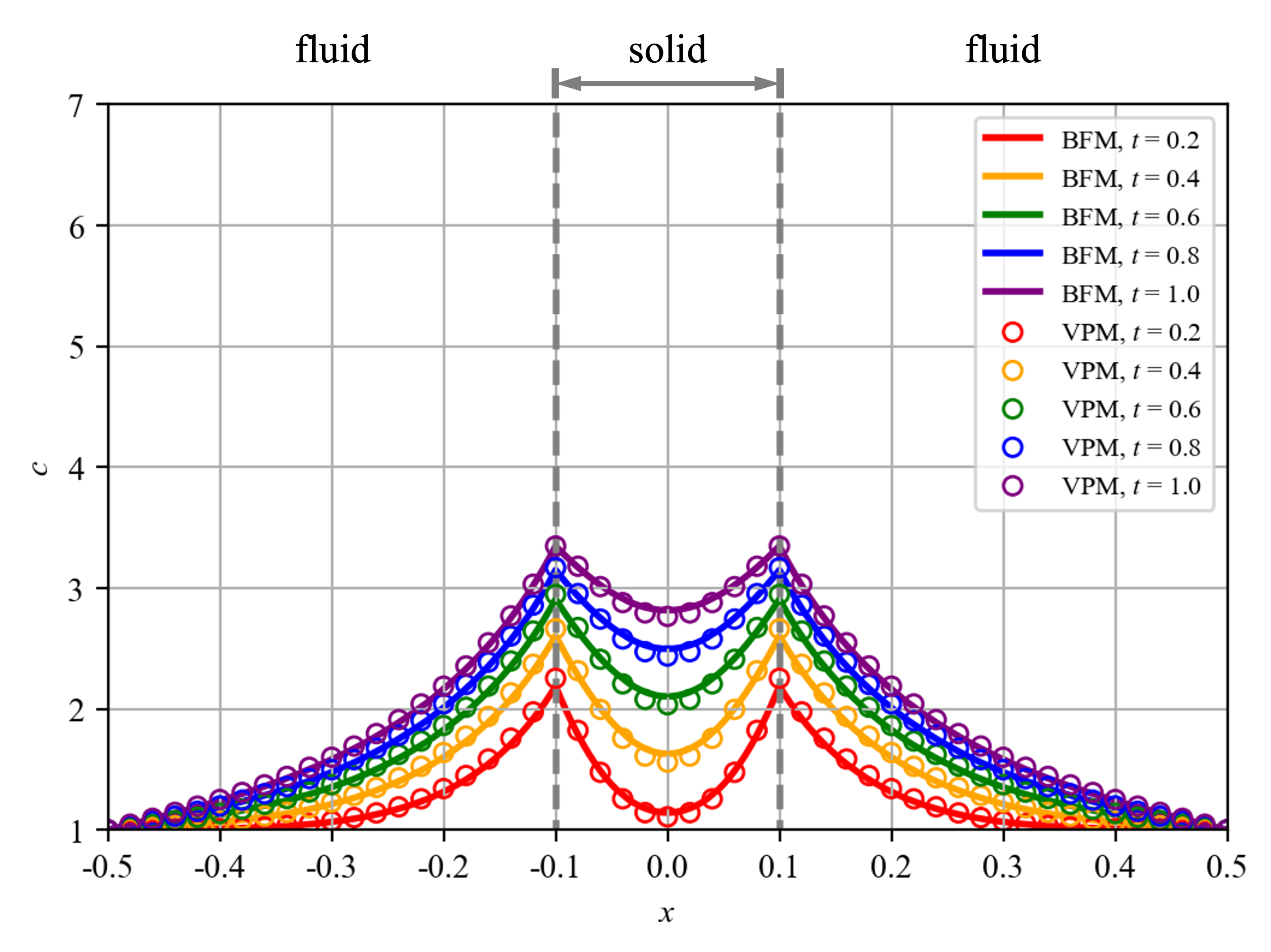}
\end{subfigure}

\caption{Distribution of the scalar along the line $\mathit{y}=0$ for (a) Case 1 and (b) Case 2.}
\label{fig13}
\end{figure}

\begin{figure}[H]\ContinuedFloat
\centering

\begin{subfigure}{\linewidth}
  \centering
  \caption{}
  \includegraphics[width=0.8\linewidth]{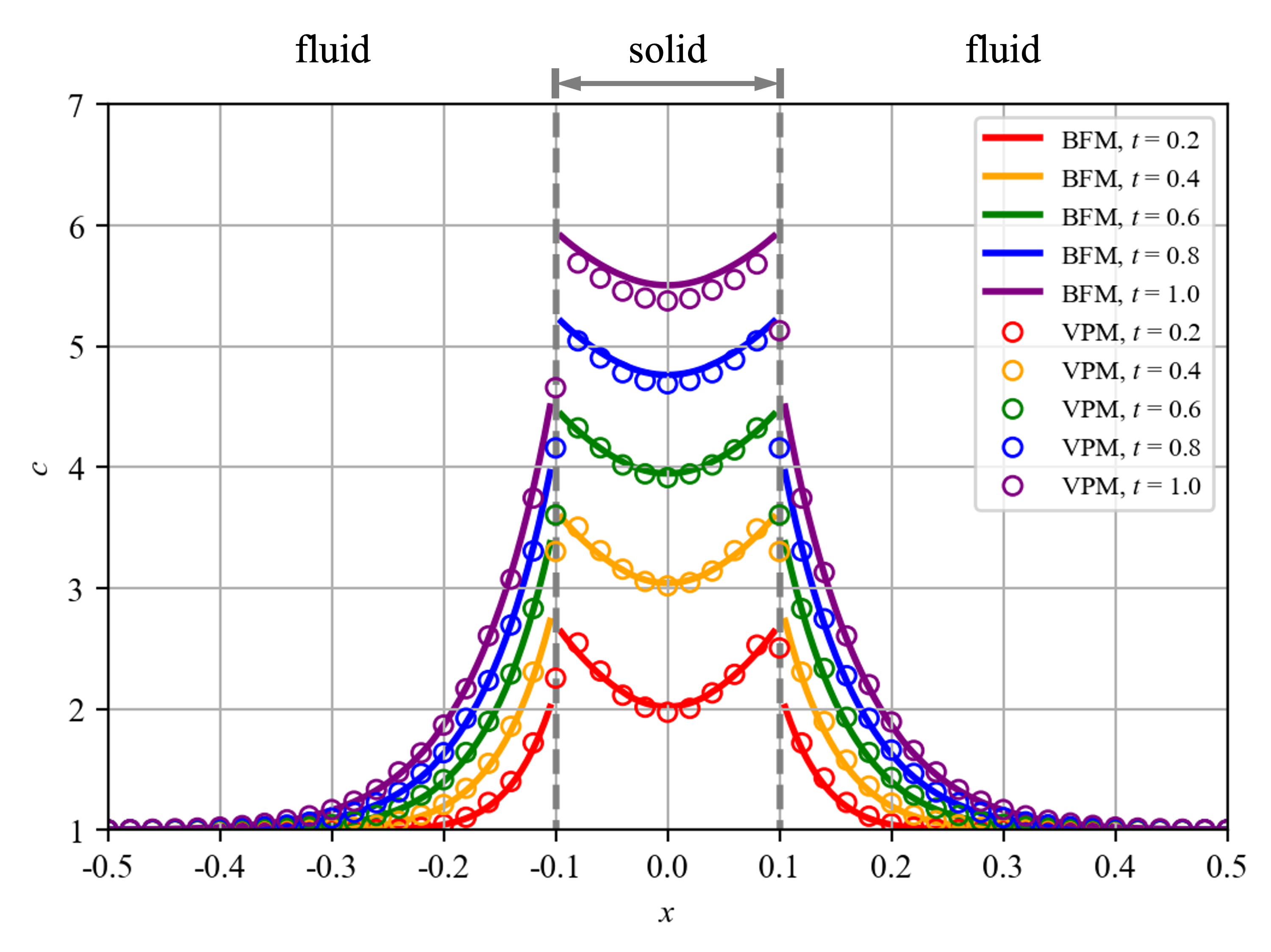}
\end{subfigure}

\medskip

\begin{subfigure}{\linewidth}
  \centering
  \caption{}
  \includegraphics[width=0.8\linewidth]{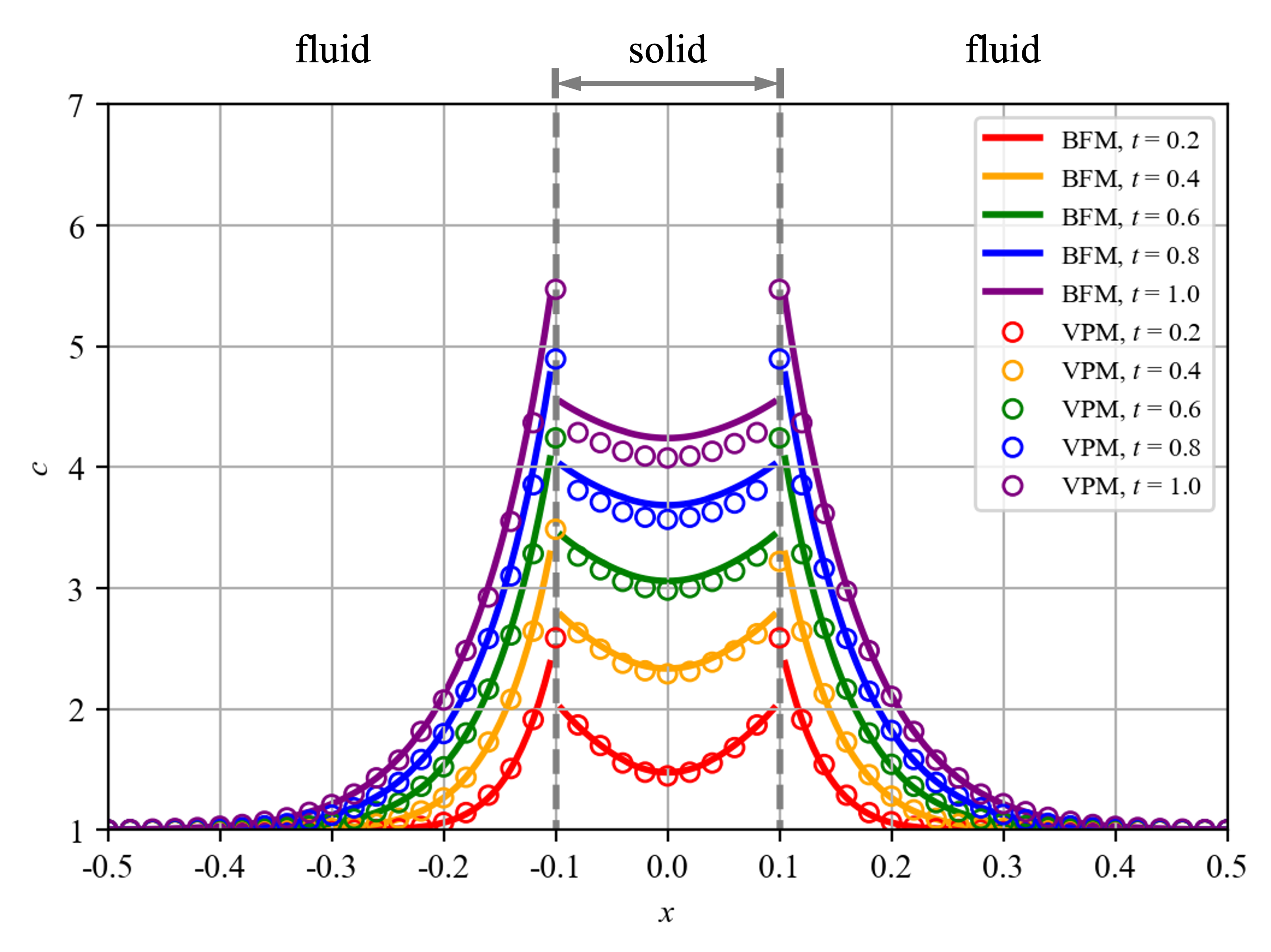}
\end{subfigure}

\caption{Distribution of the scalar along the line $\mathit{y}=0$ for (c) Case 3 and (d) Case 4.}
\end{figure}

\begin{figure}[H]
\centering
\includegraphics[width=0.8\textwidth]{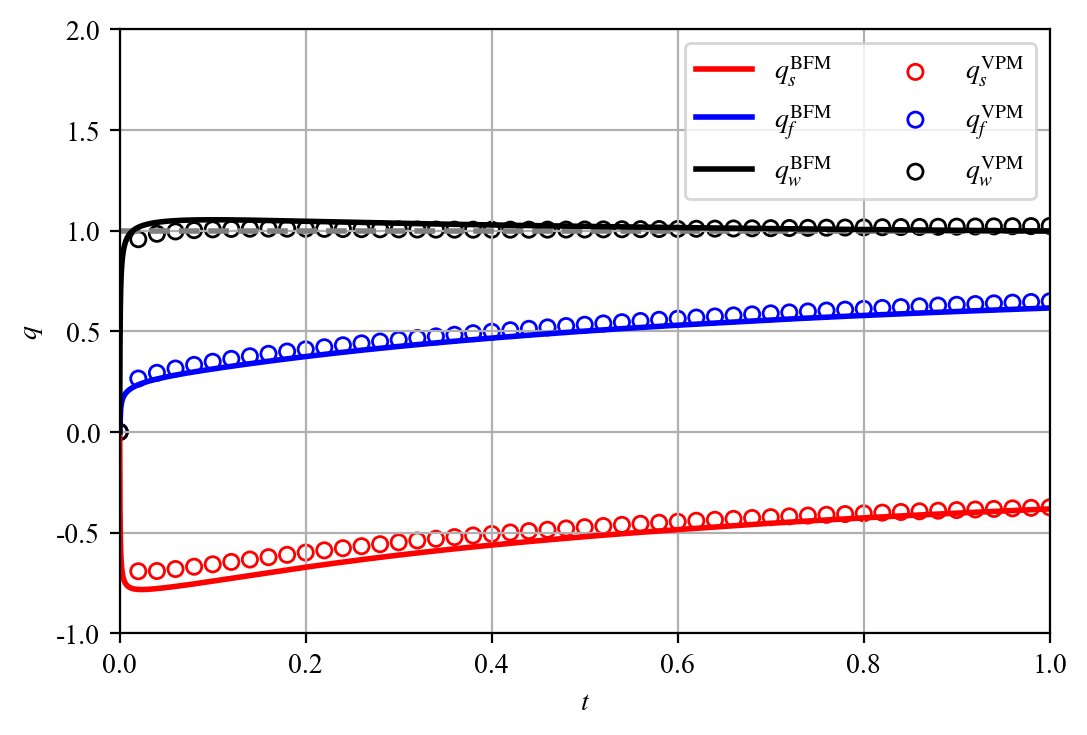}
\caption{Time evolution of the scalar flux on the solid side, the scalar flux on the fluid side, and the flux jump at the interface for Case 4.}\label{fig14}
\end{figure}

Moreover, since the present VPM is originally proposed to treat the Neumann boundary condition, the scalar flux across the interface is analyzed here by taking Case 4 as an example. Figure 14 plots the time evolution of the simulated scalar flux around the interface $\mathit{x}$ = 0.1, $\mathit{y}$ = 0, i.e., the scalar flux on the solid side $\mathit{q}_{s}$, the scalar flux on the fluid side $\mathit{q}_{f}$, and the flux jump at the interface $\mathit{q_w}$, respectively. They are respectively defined as follows,

\begin{equation}
q_s^{\mathrm{VPM}}
=
- D_s
\left.
\frac{\partial c^{\mathrm{VPM}}}{\partial n}
\right|_{\phi_0=\delta} ,
\end{equation}

\begin{equation}
q_s^{\mathrm{BFM}}
=
- D_s
\left.
\frac{\partial c_s^{\mathrm{BFM}}}{\partial n}
\right|_{\phi_0 = 0^{+}} ,
\end{equation}

\begin{equation}
q_s^{\mathrm{VPM}}
=
- D_s
\left.
\frac{\partial c^{\mathrm{VPM}}}{\partial n}
\right|_{\phi_0=\delta} ,
\end{equation}

\begin{equation}
q_f^{\mathrm{BFM}}
=
- D_f
\left.
\frac{\partial c_f^{\mathrm{BFM}}}{\partial n}
\right|_{\phi_0 = 0^{-}} ,
\end{equation}

\begin{equation}
q_w^{\mathrm{VPM}}
=
q_f^{\mathrm{VPM}} - q_s^{\mathrm{VPM}}
=
- D_f
\left.
\frac{\partial c^{\mathrm{VPM}}}{\partial n}
\right|_{\phi_0 = -\delta}
+
D_s
\left.
\frac{\partial c^{\mathrm{VPM}}}{\partial n}
\right|_{\phi_0 = \delta} ,
\end{equation}

\begin{equation}
q_w^{\mathrm{BFM}}
=
q_f^{\mathrm{BFM}} - q_s^{\mathrm{BFM}}
=
- D_f
\left.
\frac{\partial c_f^{\mathrm{BFM}}}{\partial n}
\right|_{\phi_0 = 0^{-}}
+
D_s
\left.
\frac{\partial c_s^{\mathrm{BFM}}}{\partial n}
\right|_{\phi_0 = 0^{+}} .
\end{equation}
where $\phi_0 = \delta$ and $\phi_0 = -\delta$  correspond to the solid-side and fluid-side boundaries of the interfacial region ($-\delta < \phi_0 < \delta$), respectively, as shown in Fig. \ref{fig3}(a). As shown in Fig. \ref{fig14}, the temporal evolution of the scalar fluxes around the interface can be produced accurately using the present VPM.

\section{Fluid-solid coupled advection-diffusion problem}
\label{sec5}
\subsection{Problem description}
\label{subsec5_1}
\begin{figure}[H]
\centering
\includegraphics[width=\textwidth]{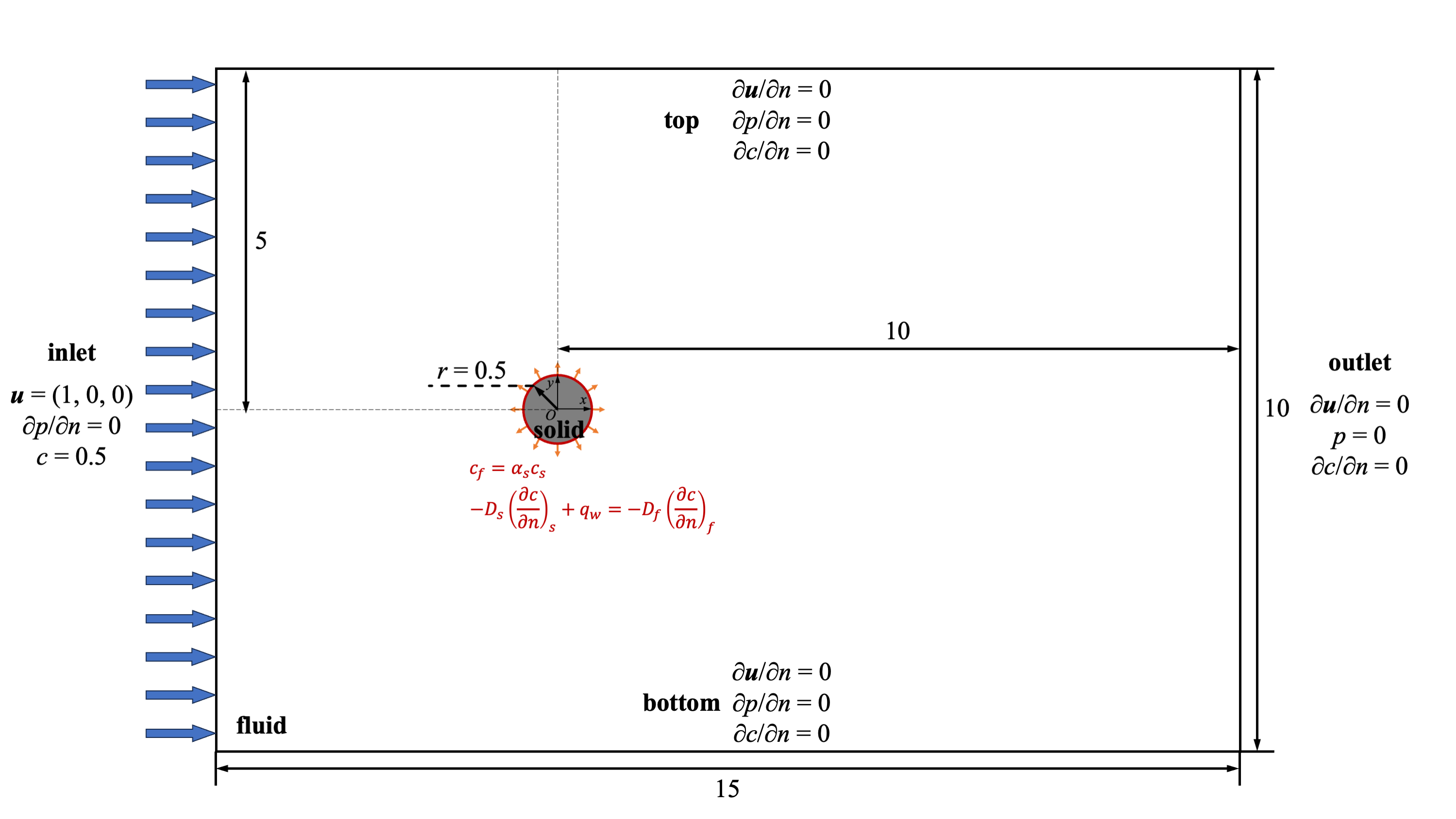}
\caption{Schematic of fluid-solid coupled advection-diffusion problem. Here, the white and grey regions represent fluid and solid phases, respectively, and the red circle denotes the fluid-solid interface. The orange arrows show the uniform distribution of constant scalar flux $\mathit{q_w}$ on the interface.}\label{fig15}
\end{figure}

In this section, the present VPM is further extended to a two-dimensional fluid-solid coupled advection diffusion problem. As illustrated in Fig. \ref{fig15}, the steady incompressible laminar flow past a circular cylinder with a radius $\mathit{r}$ = 0.5 is considered in the computational domain with the size of [-5, 10]$\times$[-5, 5]. The origin of the coordinates is defined at the center of the cylinder. The cylinder diameter $\mathit{D}_\mathrm{cylinder}$ and the inflow velocity $\mathit{U}_\mathrm{in}$ are used as characteristic length- and velocity-scales, resulting in the Reynolds number of $\mathit{Re} = \frac{\mathit{U}_\mathrm{in} \mathit{D}_\mathrm{cylinder}}{\nu}$ = 40, where $\nu$ is the kinematic viscosity. The present condition corresponds to a steady flow regime. We consider a constant scalar flux jump $\mathit{q_w}$ at the fluid-solid interface. The boundary conditions for the fluid flow consist of a fixed value for the velocity at the inlet, a fixed value for the pressure at the outlet, and zero gradients at the top and bottom walls. As for the boundary conditions for the scalar transport, a fixed value of 0.5 is given at the inlet, and zero gradient boundary conditions are imposed at the outlet, top, and bottom boundaries. The governing equations in each phase and the boundary condition at the fluid-solid interface are written as

\begin{equation}
\left\{
\begin{aligned}
\nabla \cdot \boldsymbol{u} &= 0, 
&& \text{in } \Omega_f, \\
\boldsymbol{u} \cdot \nabla \boldsymbol{u}
&= - \nabla p + \frac{1}{Re} \nabla^{2} \boldsymbol{u},
&& \text{in } \Omega_f, \\
\boldsymbol{u} \cdot \nabla c_f
- \frac{1}{Re Sc} \nabla^{2} c_f
&= 0,
&& \text{in } \Omega_f, \\
- \frac{\kappa}{Re Sc} \nabla^{2} c_s
&= 0,
&& \text{in } \Omega_s ,
\end{aligned}
\right.
\end{equation}

\begin{equation}
\left\{
\begin{aligned}
c_f &= \alpha_s c_s, \\
- \frac{\kappa}{Re Sc}
\left. \frac{\partial c}{\partial n} \right|_{s}
+ q_w
&=
- \frac{1}{Re Sc}
\left. \frac{\partial c}{\partial n} \right|_{f},
\end{aligned}
\right.
\qquad \text{on } \partial \Omega_{fs} ,
\end{equation}
where the Schmit number is defined as $\mathit{Sc} = \frac{\nu}{\mathit{D}_{f}}$ and $\kappa = \frac{D_s}{D_f}$ is the ratio of the molecular diffusivity between solid and fluid. The unified VPM governing equations for both the fluid and solid phases are given by

\begin{equation}
\left\{
\begin{aligned}
\nabla \cdot \boldsymbol{u} &= 0, \\
\boldsymbol{u} \cdot \nabla \boldsymbol{u}
&= - \nabla p
+ \frac{1}{\mathrm{Re}} \nabla^{2} \boldsymbol{u}
- \eta\, \phi \, \boldsymbol{u}, \\
\boldsymbol{u} \cdot \nabla h
&=
\frac{1}{\mathrm{Re}\,\mathrm{Sc}} \nabla \cdot ( D \nabla h )
+ \nabla \cdot \bigl( q_w^{*} \, \phi \, \boldsymbol{n} \bigr) ,
\end{aligned}
\right.
\end{equation}
where
\begin{equation}
D(\phi_0)
=
\begin{cases}
1, & \phi_0 < 0, \\
\kappa, & \phi_0 \ge 0 .
\end{cases}
\end{equation}
In the present case, the constant parameters are set as $\mathit{Re}$ =40, $\mathit{Sc}$ = 0.5, $\kappa$ = 5, $\alpha_s$ = 1.25, and $\mathit{q_w}$ = 100.0.

\begin{figure}[H]
\centering
\includegraphics[width=\textwidth]{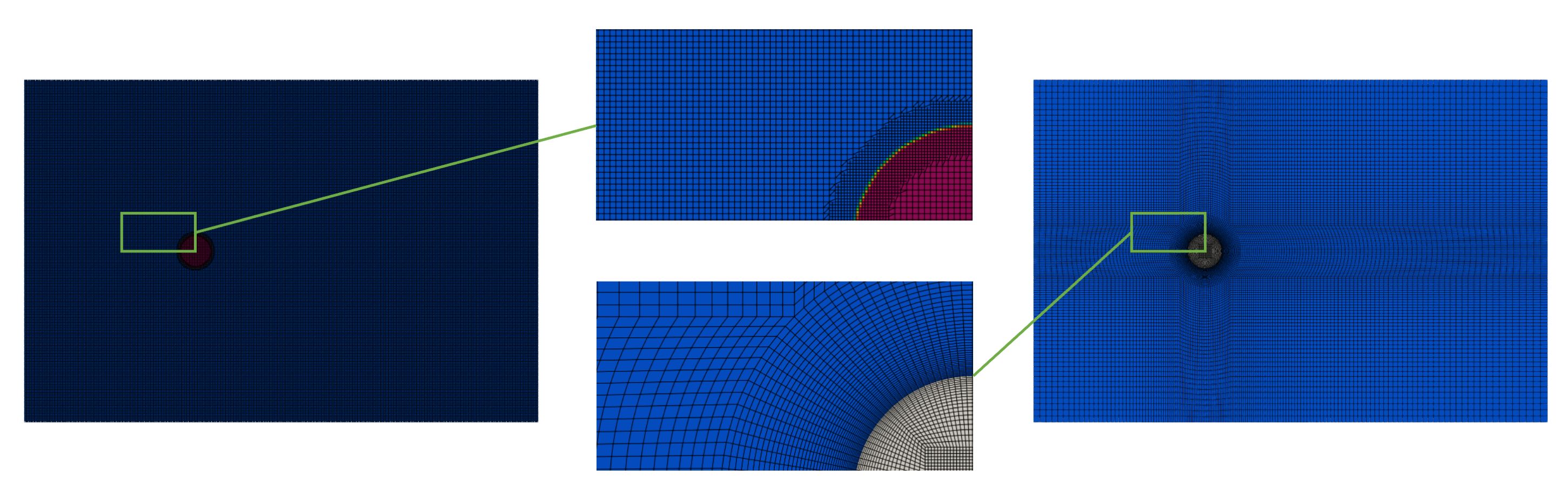}
\caption{Grids distribution of Cartesian grids (left) and body-fitted grids (right) for the fluid-solid coupled advection-diffusion problem.}\label{fig16}
\end{figure}

Similar to the fluid-solid coupled scalar diffusion problem in Sec. \ref{sec4}, the Cartesian grids with the grid size of 0.005 are used for the present VPM. Then, the VPM results are also compared with the BFM results simulated by chtMultiRegionFoam in OpenFOAM. The distributions of Cartesian grids and body-fitted grids are displayed in Fig. \ref{fig16}.

\subsection{Results and discussions}
\label{subsec5_2}
\begin{figure}[H]
  \centering
  \begin{subfigure}{\linewidth}
    \caption{}
    \centering
    \includegraphics[width=0.8\linewidth]{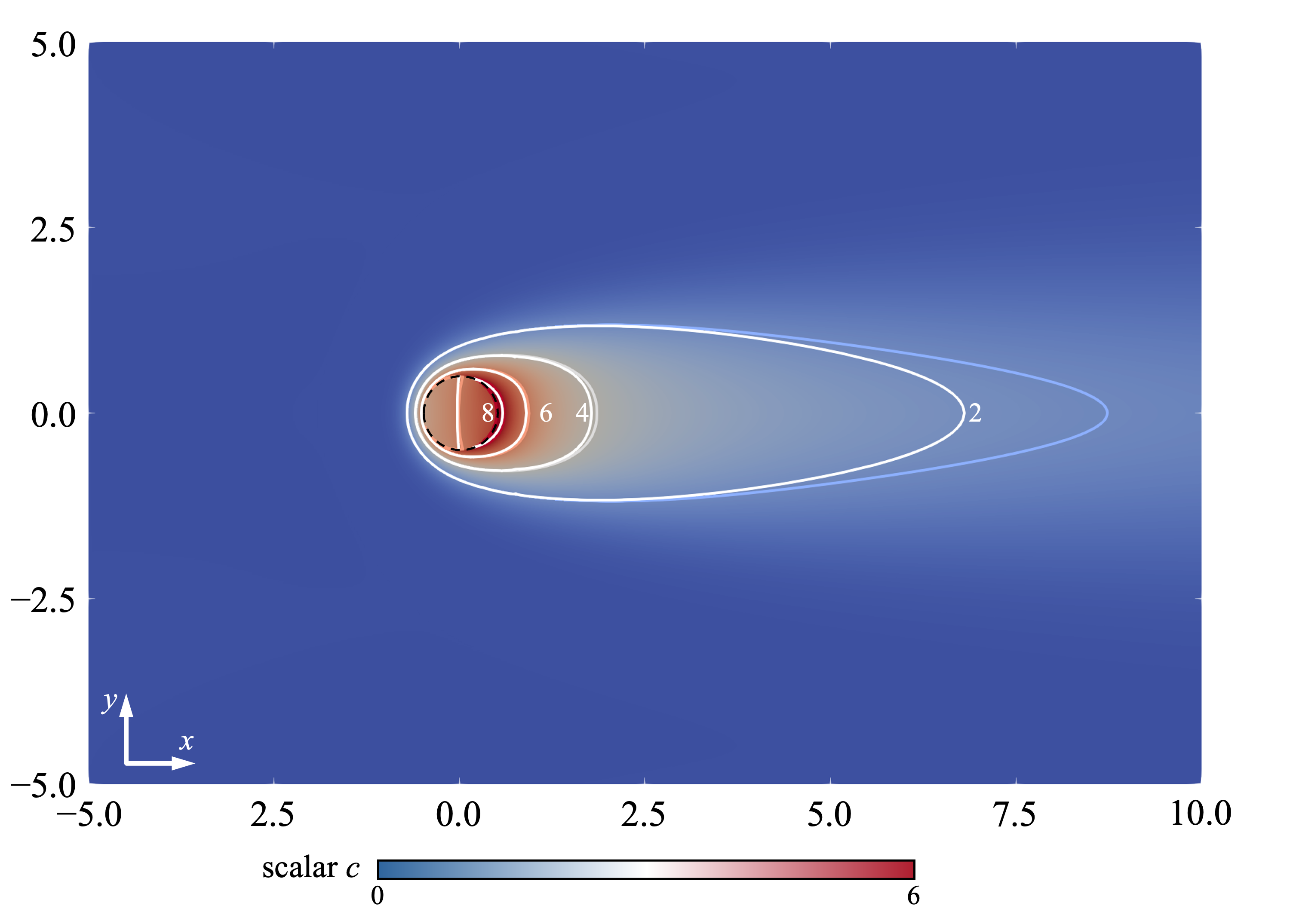}
  \end{subfigure}\par\medskip
  \begin{subfigure}{\linewidth}
    \caption{}
    \centering
    \includegraphics[width=0.8\linewidth]{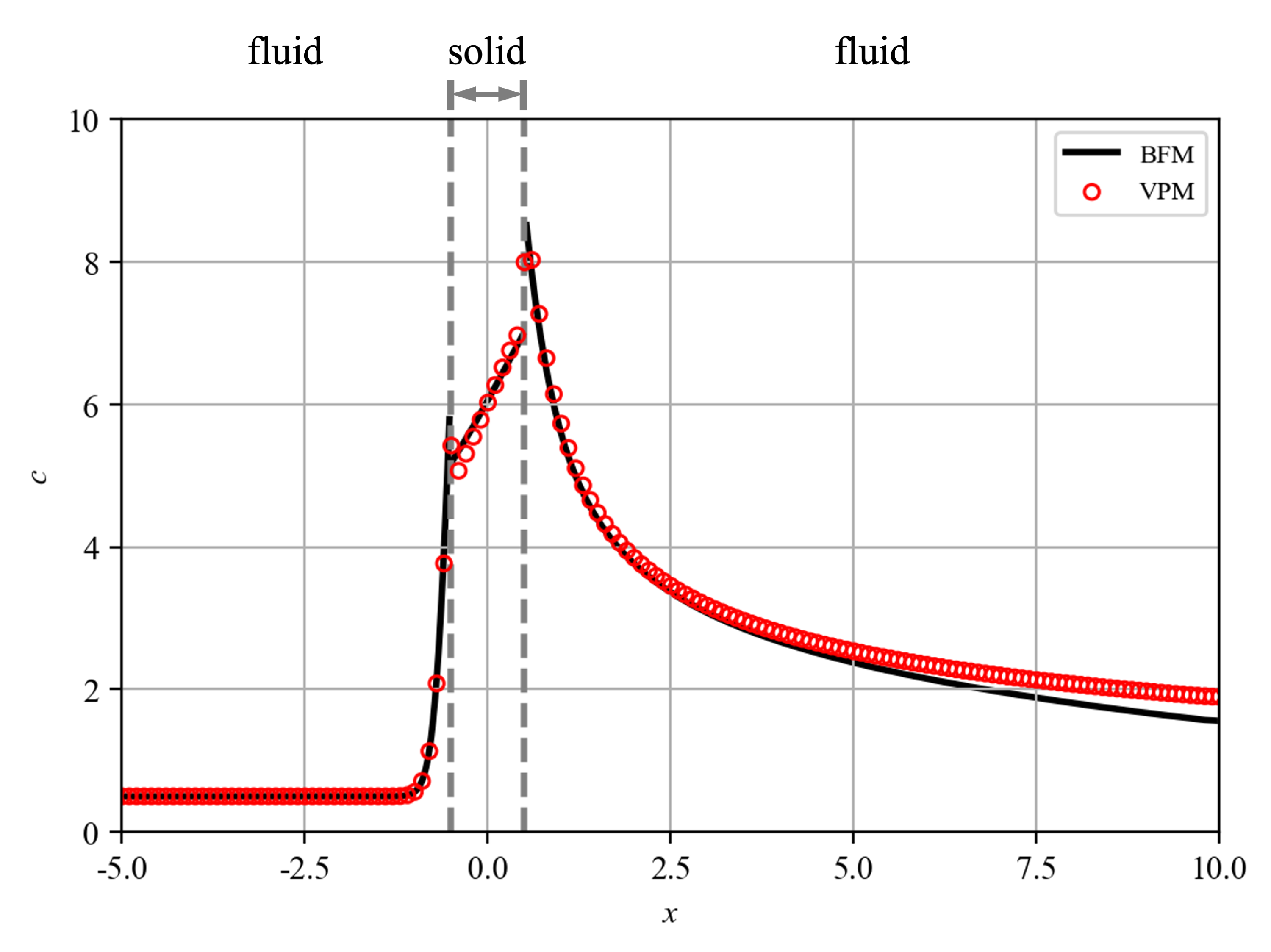}
  \end{subfigure}\par\medskip
  \caption{Distribution of the steady-state scalars (a) inside the domain and (b) along the line $\mathit{y}$ = 0. In Fig. 17(a), the contours with colored isolines are plotted by VPM results, and the white isolines are plotted by BFM results. The black dashed circle indicates the fluid-solid interface.}
  \label{fig17}
\end{figure}

Figure \ref{fig17} shows the contours of the steady-state scalar as well as their distributions along the line y = 0 obtained by the present VPM and BFM solvers. From the comparison, it can be confirmed that the present VPM results agree well with those obtained by the body-fitted mesh. The interfacial jump of the scalar and its flux near the interface can be reproduced accurately by the present method. Quantitative comparison between VPM and BFM results shows that their relative deviation defined by Eq. \ref{eq42} is 1.59\%. The result validates that the present VPM also works for the advection-diffusion conjugate transport problem with the interfacial jump condition.

\section{Conclusions}
\label{sec6}

A volume penalization method, a kind of diffuse immersed boundary method, has been widely used in numerical simulations of flow and heat transfer with complex geometries to avoid cumbersome grid generation. Despite its great success in implementing Dirichlet boundary conditions, attempts to apply it to Neumann boundary conditions are still limited. In the conventional VPM that forces the derivative of a scalar to a target value inside the fluid region, an artificial scalar field is also diffused inside the solid. As a result, additional procedures are always required to deal with conjugate transport problems coupling fluid and solid phases.

In the present study, a VPM for solving conjugate scalar transport with interfacial jump conditions is newly proposed and validated. First, a novel VPM where a source term has a divergence form is proposed to establish a unified advection-diffusion equation with the Neumann boundary condition. Then, by introducing an equivalent scalar as well as an equivalent flux, the proposed VPM is further extended to solve conjugate scalar transport problems with the jump in scalar and its flux.

A one-dimensional diffusion problem is employed to verify the accuracy of the proposed VPM. The first-order accuracy is confirmed by grid convergence studies, and the relative error of the present VPM to analytical solutions is around 1\% on average. The comparison between the present VPM and the existing scheme [26] shows that the present scheme can not only achieve higher simulation accuracy but also avoid the generation of a non-physical scalar profile inside the solid region.

Fluid-solid coupled diffusion and advection-diffusion problems with jumps in scalar and its flux are then simulated by the present scheme. The simulation results are compared with those using BFM, and the relative deviations are within 3.0\%, validating the present VPM for treating interfacial jump conditions. 

The unique feature of the present VPM lies in that both fluid and solid regions are simulated by solving a unified governing equation. It has a great advantage in dealing with complex geometries, e.g., rough walls \cite{Thakker18}, and complex heat transfer surfaces \cite{Kametani20}. Recently, VPMs have been developed for coupled radiation, convection, and conduction problems \cite{LiuM23}\cite{ChenD25}. The present VPM for interfacial jump conditions can easily be extended to more complex forward problems mentioned above. In addition, VPMs can be combined with the adjoint method for shape and topology optimization in various flow and heat transfer problems, such as radiative transfer \cite{LiuM24}, turbulent heat transfer \cite{Kametani20}. and conjugate heat transfer \cite{ChenD25}. Applications of the present VPM to them are quite promising and should be tackled in future work.

\section*{Acknowledgments}
The present research is partially supported by Research and Development Program for Promoting Innovative Clean Energy Technologies Through International Collaboration, the New Energy and Industrial Technology Development Organization (NEDO), Japan; and International Research Fellow of Japan Society for the Promotion of Science (Postdoctoral Fellowships for Research in Japan (Standard)), Japan Society for the Promotion of Science, Japan. Y.H. also gratefully acknowledges the supports from JSPS KAKENHI Grant Number JP23K26034, and also Adaptable and Seamless Technology Transfer Program through Target-driven R\&D (A-STEP) from Japan Science and Technology Agency (JST) Grant Number JPMJTR232D.

\section*{Declaration of interests}
The authors declare that they have no known competing financial interests or personal relationships that could have appeared to influence the work reported in this paper.

\end{document}